\documentclass[11pt]{article}
\pdfoutput=1 
\usepackage{jheppub}
\usepackage{graphicx,verbatim, upgreek, txfonts, yfonts}
\usepackage{psfrag, times}
\usepackage[T1]{fontenc}
\usepackage{extarrows}

\def\be{\begin{equation}}
\def\ee{\end{equation}}
\def\bc{\begin{center}}
\def\ec{\end{center}}

\def\bG{{\bf{G}}}

\def\by{{\mathbf{y}}}

\def\cF{{\mathcal{F}}}

\def\cN{{\mathcal{N}}}

\def\cW{{\mathcal{W}}}

\def\nn{\nonumber}

\def\q{{\bf q}}

\def\r2{{\sqrt{2}}}

\def\z{{\zeta}}

\def\be{\begin{equation}}
\def\ee{\end{equation}}
\def\bea{\begin{eqnarray}}
\def\eea{\end{eqnarray}}
\def\nn{\nonumber}

\def\Q{\hat {Q}}
\def\m{\hat{m}}
\def\z{\hat{\zeta}}
\def\cN{{\mathcal{N}}}
\def\a{\alpha}
\def\b{\beta}
\def\({\left(}
\def\){\right)}
\def\0.5{\frac{1}{2}}
\def\bs{\begin{split}}
\def\es{\end{split}}
\def\pl{\prod\limits}
\def\m{\tilde{m}}
\def\q{\tilde{q}}
\def\s{\tilde{s}}
\def\l||{\left|\left|}
\def\r||{\right|\right|}
\def\q{\tilde{q}}

\def\mi2{\frac{m_i}{2}}
\def\mj2{\frac{m_j}{2}}
\def\fB{\mathfrak{B}}
\def\a{\alpha}
\def\b{\beta}
\def\qq{q^{\frac{1}{2}}}
\def\phii{\tilde{\phi}}

\def\hf{\hat{f}}
\def\hg{\hat{g}}

\begin{document}

\title{\LARGE \bf Connecting Mirror Symmetry in 3d and 2d via Localization}

\author{ Heng-Yu Chen, Hsiao-Yi Chen and Jun-Kai Ho}
\affiliation{ Department of Physics and Center for Theoretical Sciences, \\
National Taiwan University, Taipei 10617, Taiwan}
\emailAdd{ heng.yu.chen@phys.ntu.edu.tw} 
\emailAdd{b98202059@ntu.edu.tw}
\emailAdd{b98202064@ntu.edu.tw}
\vspace{2cm}

\abstract{We explicitly apply localization results to study the interpolation between three and two dimensional mirror symmetry for Abelian gauge theories with four supercharges. We first use the ellipsoid $S_b^3$ partition functions to verify the mirror symmetry between a pair of general three dimensional ${\mathcal{N}}=2$ Abelian Chern-Simons quiver gauge theories. 
These expressions readily factorize into holomorphic blocks and their anti-holomorphic copies, so we can also obtain the partition functions on $S^1\times S^2$ via fusion procedure. We then demonstrate $S^1\times S^2$ partition functions for the three dimensional Abelian gauge theories can be dimensionally reduced to the $S^2$ partition functions of ${\mathcal N}=(2,2)$ GLSM and Landau-Ginzburg model for the corresponding two dimensional mirror pair, as anticipated previously in \cite{Aganagic:2001uw}. 
We also comment on the analogous interpolation for the non-Abelian gauge theories and compute the K-theory vortex partition function for a simple limit to verify the prediction from holomorphic block. }

\maketitle

\section{Introduction}
\paragraph{}
Recently, there have been resurgent interests in studying various infra-red dualities for supersymmetric gauge theories in different dimensions. Typically these dualities involve two seemingly different theories at high energy, but they exhibit exactly the same infra-red behavior.
In three dimensions, perhaps the best known example of such dualities is the Coulomb-Higgs duality or so-called ``3d mirror symmetry''  for wide range of $\cN=4$ abelian and non-abelian gauge theories discovered in \cite{Intriligator:1996ex}, and later these examples were also generalized to $\cN=2$ gauge theories \cite{Aharony:1997bx, deBoer:1997ka, Dorey:1999rb} through various addition of superpotential terms or gauging of global symmetries. 
However, unlike in four dimensions, abelian gauge theories also become strongly coupled at low energy in three dimensions, which restricted the early verifications of the three dimensional mirror pairs to such as merely matching the global parameters, global symmetries and dimensions of the moduli spaces, and many examples directly invoked the $SL(2,{\mathbb Z})$ duality in type-IIB string theory following \cite{Hanany:1996ie}.
While, in two dimensions, there is also the famous original mirror symmetry relating $\cN=(2,2)$ abelian gauged linear sigma models and Landau-Ginzburg models with Toda-type superpotentials, which was later interpreted as generalization of T-duality transformation in \cite{Hori:2000kt, Rocek:1991ps}.  However, for more general non-abelian gauged linear sigma models which are associated with Grassmannian or flag manifolds, the abelian duality transformation can no longer be applied to derive the mirror dual geometries.

The striking new developments in computing the exact partition functions of supersymmetric gauge theories defined on different curved manifolds using localization techniques provide us new tools to test these dualities \cite{Kapustin:2009kz, Kapustin:2010xq, Hama:2010av, Hama:2011ea, Benvenuti:2011ga, Imamura:2011wg, Benini:2012ui, Doroud:2012xw, Gomis:2012wy} and also allow us to interpolate partition functions between different dimensions \cite{Dolan:2011rp, Gadde:2011ia, Imamura:2011uw}. 
These lead us to the natural question that is whether the non-trivial infra-red dualities in different dimensions can be related via dimensional reduction. 
The answer seems rather difficult in general as compactification typically probes the ultraviolet behavior, and indeed as discussed in \cite{Aharony:2013dha, Aharony:2013kma} for correct interpolation between four and three dimensional non-abelian dualities \cite{Seiberg:1994pq, Aharony:1997gp},
additional non-perturbatively generated superpotentials need to be included. 
In this paper, we are interested in verification and interpolation of three and two dimensional mirror symmetry, utilizing the recent localization results. 
This represents our initial step in understanding the relations among dualities in three and two dimensions, in particular focusing on the general three dimensional $\cN=2$ abelian mirror pair constructed in \cite{Dorey:1999rb, Tong:2000ky}, derived from three dimensional $\cN=4$ abelian mirror pair in \cite{Kapustin:1999ha}.
We first explicitly match their partition functions computed on ellipsoid $S_b^3$, which non-trivially verifies the three dimensional mirror symmetry for this class of theories. Interestingly, these partitions exhibit beautiful factorization property after we appropriately interpret different vacua \cite{Pasquetti:2011fj, Beem:2012mb}, and this property allows us to also obtain the partition function on $S^1 \times_q S^2$ by fusion procedure. For finite but small $S^1$ radius, we can rewrite the integrand into exponential of holomorphic and anti-holomorphic copies of the twisted superpotential obtained from integrating out chiral fields and its KK modes, as noticed in \cite{Aganagic:2001uw}. 
In fact, for arbitrary $S^1$ radius, we can rewrite the integrand in terms of the q-deformed generalization of the Fourier transform for Bessel functions.  

Finally, taking either one of the three dimensional mirror pair theories, we explicitly reduce along the $S^1$ fibre of its $S^1\times_q S^2$ partition function, and demonstrate that after regularization to include the KK-modes,  the partition function can either be written as the $S^2$ partition for a $\cN=(2,2)$ GLSM or Landau-Ginzburg model as computed in \cite{Benini:2012ui, Doroud:2012xw, Gomis:2012wy}. From the matching of the three dimensional partition functions, we can now trivially match the $S^2$ partition function for $\cN=(2,2)$ GLSM derived from one side of three dimensional mirror pair with the $\cN=(2,2)$ Landau-Ginzburg model derived from the other side, establishing the interpolation between three and two dimensional mirror symmetry dualities. 
In addition, in the last section, we explicitly demonstrate the existence of Higgs branch fixed point and calculate the K-theoretic vortex partition function in a simple case to verify the factorized gauge theory partition function. We also propose how more general cases can be reproduced.

\section{Review on Mirror Symmetry in 2+1 and 1+1 Dimensions}\label{ReviewMirror}

\subsection{Review on 3d Supersymmetric Gauge Theories}\label{3dReview}
\paragraph{}
Let us begin reviewing some necessary details about 3d $\mathcal{N}=2$ supersymmetry. We shall follow the conventions in \cite{Aharony:1997bx}, and this will help us to set up our subsequent notations.
$\mathcal{N}=2$ supersymmetry in three dimension has four real supercharges which can be obtained from dimensional reduction of four dimensional $\mathcal{N}=1$ supersymmetry:
\be
\{Q_\alpha ,Q_\beta\}=\{\bar{Q}_\alpha ,\bar{Q}_\beta  \}=0,
\quad
\{Q_\alpha , \bar{Q}_\beta\}=2\sigma^\mu _{\alpha\beta}P_\mu +2i\epsilon_{\alpha\beta}Z .
\ee
Supercharges $Q$ and $\bar{Q}$ are complex and have a $U(1)_R$ R-symmetry. The matrices $\sigma^\mu$ can be taken real and symmetric while the real central charge $Z$ comes from the momentum $P_3$ along the compactified direction.  

We will mostly focus on abelian quiver gauge theories,  whose gauge field is contained in the vector multiplet $V_a$, $a=1,2\cdots L$, satisfying $V_a=V_a^\dagger$;
and we will further couple these theories with matters given by chiral multiplet $X_i$ such $\bar{D}_\alpha X_i=0$. 
Instead of using the usual field strength multiplet as in four dimensions, in three dimensions we can alternatively introduce linear multiplet, $\bG_a=\epsilon^{\alpha\beta}\bar{D}_\alpha D_\beta V_a$,  satisfying $D^2\bG_a=\bar{D}^2\bG_a=0$.
The lowest component of $\bG_a$  is precisely the real scalar in the vector multiplet $V_a$,
and its expansion also contains a vector field term built from gauge field strength $\bar{\theta} \sigma_\rho \theta F_{\mu\nu (a)}\epsilon^{\rho\mu\nu} $.
The total kinetic terms can now be combined into a D-term
\be
\mathcal{L}_{\rm K} =\int d^4 \theta
\left[
\sum\limits^N_{i=1} X^\dagger_i\right. \exp\left(
2\sum\limits^s_{a=1}Q^a_iV_a
+2m_i \theta\theta^\dagger
\right) \left. X_i
+\sum\limits^L_{a=1}\frac{1}{e^2_a}\bG^2_a
\right],
\ee
where the $Q^a_i$ is the charge for each parameter chiral multiplet $X_i$ under $U(1)_a$, and the $e_a$ is the coupling constant for $U(1)_a$ vector field with mass dimension $\frac{1}{2}$, which implies theory can become strongly coupled at infra-red, even for Abelian gauge group.
Here we have also included the real mass $m_i$ which corresponds to the vev of the scalar component in the background vector superfield for weakly gauged flavor symmetry.
As we will discuss more later, only $N-L$ out of the $N$ real mass parameters are physical as $L$ of them can be absorbed into the lowest component of the linear multiplets under global part of each $U(1)$ gauge transformation. 
We can also write down a superpotential for interactions among chiral multiplets:
\be
\mathcal{L}_{\rm F}=\int d^2\theta\mathcal{W}\left( X_i\right) +h.c.
\ee
where $h.c.$ denotes the hermitian conjugate. 
The usual complex mass terms for chiral multiplets can be introduced through superpotential. We will however mostly consider the situation where $\cW=0$ unless otherwise stated.

In contrast with four dimensions, we can also include Chern-Simons terms for the vector superfields $\{V_a\}$ in three dimensions: 
\be
\mathcal{L}_{\rm CS}=\sum_{a,b=1}^L \kappa^{ab}_{\rm eff} \int d^4\theta \bG_a V_b. 
\ee
The effective Chern-Simons coefficient $\kappa^{ab}_{\rm eff}$ is required to be integer valued to ensure the gauge invariance.
It consists of both bare classical contribution which can induce parity anomaly and quantum mechanical one-loop contribution coming from integrating out the massive auxiliary fermions charged under the gauge group, i.e.
\be\label{CSeff}
\kappa^{ab}_{\rm eff} = \kappa^{ab}+\frac{1}{2}\sum_{i=1}^{N} Q^a_{i}Q^b_{i} {\rm sign} (M_i)
\ee 
where $\{M_i\}$ are the real effective masses of the auxiliary fermions integrated out, and in present context of vanishing superpotential, $M_i=\sum_{a=1}^s Q_{i}^a \sigma_a+m_i$ and $\sigma_a$ is the real scalar of the $U(1)_a$ vector multiplet.
Furthermore by similar mechanism, we can generate the mixed Chern-Simons terms or BF couplings between the global and gauge symmetries when we weakly gauge the background gauge field associated with the global symmetries. They can be expressed as:
 \be
\mathcal{L}_{\rm Mix}=\kappa^{a}_i \int d^4\theta \bG_a V^i = \kappa^{a}_i\int d^4\theta \bG^i V_a,
\ee
where coefficient $\kappa_{i}^a=\frac{1}{2}Q^a_i  {\rm sign} (M_i)$. 
As usual, for gauge group containing $U(1)$ factors, we can introduce one bare FI parameter multiplying with each vector super field $V_a$.
However, from our previous discussion on Chern-Simons terms, it should be clear that  when the scalar component of the linear multiplet for gauge or global symmetries acquires non-vanishing vacuum expectation value,  they provide additional contributions depending linearly on $V_a$, and we can combine these into effective FI term $\zeta_{\rm eff}^a$:
\be
\mathcal{L}_{\rm FI}=\sum\limits^L_{a=1} \zeta^a_{\rm eff}\int d^4\theta V_a 
\ee
where
\be\label{FIeff}
\zeta^a_{\rm eff}=\zeta^a+\sum_{b=1}^{L} \kappa^{ab}_{\rm eff}\sigma_b + \sum_{i=1}^N \kappa_i^a m_{i}
= \zeta^a+\frac{1}{2}\sum_{i=1}^{N} Q^a_i M_i {\rm sign}(M_i),
\ee
which can now vary with $\sigma_a$. These can also be regarded as one loop finite renormalization to the bare FI parameters $\zeta^a$. 

The moduli space of the vacua for three dimensional $\cN=2$ gauge theory can be divided into several branches. 
The first one is the Coulomb branch, where all the vacuum expectation values (vevs) of chiral multiplets vanish,
and generically the vector multiplet real scalars $\sigma_a$ take the values in the maximal tori of the original gauge group which in our case is simply $U(1)^L$ itself. We can further dualize the $L$ photons into $L$ periodic scalars $\{\gamma_a\}$, i. e. $\epsilon_{\mu\nu\rho}\partial^{\rho}\gamma_a = F_{\mu\nu (a)}$, where there exist $L$ global $U(1)_J$ symmetries which can shift each $\gamma_a$ by a constant. 
Together with $\{\sigma_a\}$ they form complex combinations $\{\sigma_a + i\gamma_a\}$, whose vevs parameterize the Coulomb branch which classically is given by $({\mathbb R}\times S^1)^L$, which can however receive both one-loop perturbative and non-perturbative 3d instanton or monopole corrections.
We can also weakly gauge the global $U(1)_J$, and the
vev of the lowest component of the linear multiplet containing the corresponding field strength is precisely
the FI parameter $\zeta^a$, through mixed Chern-Simons term with the vector superfield $V_a$. Classically non-vanishing bare FI parameter $\zeta^a$ and/or Chern-Simons coefficient $\kappa^{ab}$ can lift the Coulomb branch, and gauge symmetry is completely broken. However, Coulomb branch can still persist in our set up when we include the renormalization effects mentioned earlier, which can be achieved by judicially choosing $\zeta^a=\frac{1}{2}\sum_{i=1}^N Q_{i}^a m_i$, and also $\kappa^{ab}=\frac{1}{2}\sum_{i=1}^N Q_i^a Q_i^b{\rm sign}(M_i) $.

We can also have Higgs branch, where the vevs of vector multiplet scalar fields $\{\sigma_a+i\gamma_a\} $ are set to zero. To ensure the vanishing of the D-term potential modulo gauge transformation, the chiral multiplet scalar fields $\{x_i \}$ generally pick up non-trivial vevs, proportional to FI parameter.  For the Abelian quiver gauge theory discussed here, we have
\be
\sum_{i=1}Q^a_i |x_i|^2 = \zeta^a_{\rm eff}.
\ee
Modded out by the $U(1)^L$ rotation, such that gauge symmetry is generically broken on Higgs branch, the corresponding geometry, a toric manifold, is protected from the quantum corrections.
Higgs branch is lifted by the mass terms for chiral multiplets. Therefore, in our setup, full Higgs branch can only exist provided $m_i = 0$, upon which the effective FI $\zeta^{a}_{\rm eff}$ reduce to $\zeta^a$. 
In addition to these two separated branches, by tuning the parameters there can also be mixed branches where scalars in both vector and chiral multiplets have non-vanishing vevs.

\subsection{Mirror Symmetry in Three Dimensional Supersymmetric Gauge Theories}\label{3dMirrorSym}
\paragraph{}
Three dimensional mirror symmetry was first introduced for $\cN=4$ supersymmetric gauge theories in \cite{Intriligator:1996ex}, which is an infra-red symmetry 
exchanging the quantum mechanically corrected Coulomb branch of one theory with the classically protected Higgs branch of its mirror dual,  vice versa. 
Explicitly,  three dimensional $\cN=4$ theories have two commuting $SU(2)$ R-symmetries, sometimes denoted as $SU(2)_N$ and $SU(2)_R$, each of which acts non-trivially on either Coulomb or Higgs branches while leaving the other invariant. Triplets of masses and FI parameters also transform ${\bf(3,1)}$ and ${\bf(1,3)}$ respectively under $SU(2)_N\times SU(2)_R$. For a pair of $\cN=4$ mirror theories, both of their $SU(2)_N$ and $SU(2)_R$ symmetries, and triplets of masses and FI parameters are exchanged. 

Beginning with three dimensional $\cN=4$ mirror pairs, we can also extend the mirror symmetry to three dimensional $\cN=2$ theories by partially breaking the supersymmetries. 
An obvious way is to give mass to the $\cN=2$ chiral multiplet $\Psi$ in the full $\cN=4$ vector multiplet. This can be done by coupling it with another singlet field $S$ via a superpotential $\sim S\Psi$, which can be interpreted as a dynamical complex FI term \cite{Aharony:1997bx} while,
under $\cN=4$ mirror symmetry, a dynamical mass superpotential term is also generated in the dual mirror theory. 
An alternative way to derive new $\mathcal{N}=2$ mirror pairs from the $\mathcal{N}=4$ ones without introducing explicit superpotential,
which is the one to be considered here,  was introduced in \cite{Tong:2000ky}. 
The author of \cite{Tong:2000ky} considered the most general three dimensional $\cN=4$ Abelian mirror pair constructed in \cite{Kapustin:1999ha}, which are given by: 
\\\\
{\bf Theory A' :} $\mathcal{N}=4$ $U(1)^s$ gauge group with $N$ hypermultiplets with charge $Q^a_i$ under the $U(1)_a$. The theory has triplets of FI parameters $\vec{\zeta}^a$ and mass parameter $\vec{m}_i$. 
\\\\
\def\Q{\hat {Q}}
\def\m{\hat{m}}
\def\z{\hat{\zeta}}
{\bf Theory B' :} $\mathcal{N}=4$ $U(1)^{N-s}$ gauge group with $N$ hypermultiplets with charge $\Q^p_i$ under the $U(1)_p$. The theory has triplets of FI parameters $\vec{\z }^p$ and mass parameter $\vec{\m }_i$.
\\\\
The charges of hypermultiplets in the mirror pair also need to satisfy following orthogonal relation:
\begin{equation}
\label{charge relation}
\sum\limits^N_{i=1}Q^a_i\Q^p_i=0 \quad , \forall a, p .
\end{equation}
For generic non-vanishing values of masses $\vec{m}_i$ and FI parameter $\vec{\zeta}^a$, the $SU(2)_N\times SU(2)_R$ R-symmetries are broken completely. We can however rotate them such that each only contains one non-vanishing real component, i. e.  the aforementioned ``real mass'' and ``real FI-parameter" which still preserve residual $U(1)_N\times U(1)_R \subset SU(2)_N\times SU(2)_R$ R-symmetry group. We can make the same choices for the parameters in Theory B', again preserving residual $U(1)_N\times U(1)_R$ R-symmetries. $\cN=2$ mirror symmetry exchanges the roles of $U(1)_N$ and $U(1)_R$ in the dual mirror pair.

We can now break Theory A' and Theory B' to $\mathcal{N}=2$ by weakly gauging the axial R-symmetry $U(1)_{R-N}$ and $U(1)_{N-R}$ respectively, denoting the scalar components of the 
corresponding linear multiplets as $\mu$ and $-\mu$, which introduce additional contributions to the real mass for the $\mathcal{N}=2$ chiral multiplets. 
Now among the $\cN=2$ chiral multiplets in Theory A', $N$ of them have effective masses $(Q_i^a\sigma_a+m_i +\mu)$, $N$ of them have effective masses $-(Q_{i}^a\sigma_a+m_i -\mu)$, and finally $s$ neutral chiral multiplets with mass $\mu$ coming from $\cN=4$ vector multiplets. 
Now we can take the limit $\mu \rightarrow\infty$ while keeping fixed the combinations $Q_i^a \sigma_a+\frac{\mu}{2}$ and $m_i+\frac{\mu}{2}$. 
$N$ charged $\cN=2$ chiral multiplets can now be integrated out to generate a Chern-Simons term with coefficient $\frac{1}{2}\sum_{i=1}^{N} Q_i^a Q_i^b$, and $s$ neutral chiral multiplets are also integrated out. The FI parameter $\zeta^a$ also receives $\mu$-dependent finite renormalization. We can perform same integrating out procedure to Theory B', and up to redefinition of the mass and FI parameters in the resultant theories, we finally arrive at a new $\cN=2$ mirror pair constructed from $\cN=4$ one, which are of our main concerns for localization calculations and dimensional reduction in this note: 
\\\\
{\bf Theory A :} $U(1)^s$ gauge group with $N$ chiral multiplets with charge $Q^a_i$ under the $U(1)_a$ of real mass $m_i$. 
The classical Chern-Simons couplings are given by $\kappa^{ab}=\frac{1}{2}\sum^N_{i=1}Q^a_iQ^b_i$ and FI parameters $\zeta^a$.
\\\\
\def\Q{\hat {Q}}
\def\m{\hat{m}}
\def\z{\hat{\zeta}}
{\bf Theory B :} $U(1)^{N-s}$ gauge group with $N$ chiral mutiplets with charge $\Q^p_i$ under the $U(1)_p$ and real mass $\m_i$. The classical Chern-Simons couplings are given by $\hat{\kappa}^{pq}=-\frac{1}{2}\sum^N_{i=1}\Q^p_i\Q^q_i$ and the FI parameters $\z^p$. 
\\\\
As we expect three dimensional $\cN=2$ mirror symmetry to exchange the remaining mass and FI parameters among the mirror pairs,
however we should note that in Theory A we can shift the $s$ charged real scalar fields $\sigma_a$ while keep invariant the $(N+s)$ effective masses and effective FI parameters which also mix linearly with $\sigma_a$ by performing compensating shifts on $\{m_i\}$ and $\{\zeta^a\}$. This gives $s$-constraints, and only $N$ of these are independent physical parameters. Identical argument shows that also only $N$ out of $(2N-s)$ $\{\m_i\}$ and $\{\z^p\}$ in Theory B are independent.
Bearing these in mind, we can now state the mapping of the parameters in Theory A and Theory B under three dimensional Mirror symmetry \cite{Dorey:1999rb, Tong:2000ky}\footnote{We should however note that in curved background, there are additional contributions to FI term coming from background field strength of global symmetries. }:
\begin{eqnarray}
\label{MirrorMap}
&&\zeta ^a -\frac{1}{2}\sum\limits^N_{i=1}Q^a_im_i =\sum\limits^N_{i=1}Q^a_i\m_i, \nn
\\
&&\z^p +\frac{1}{2}\sum\limits^N_{i=1}\Q^p_i\m_i =  -\sum^N_{i=1}\Q_i^p m_i .
\end{eqnarray}
Notice that we are matching the classically protected Higgs branches with the quantum corrected Coulomb branches between the mirror pair, where the latter can only exist when we also integrate out the massive chiral multiplets to precisely cancel the judicially chosen classical Chern-Simons terms such that $\kappa^{ab}_{\rm eff}=\hat{\kappa}_{\rm eff}^{pq}=0$. 
We are therefore matching the residual effective FI parameters with the real mass parameters, and the vanishing of left and right sides in (\ref{MirrorMap}) ensures the existence of the Coulomb and Higgs branches respectively. 
In fact the vanishing of effective Chern-Simons terms is only possible if the vevs of the vector scalars $\{\sigma_a\}$ and $\{\hat{\sigma}_p\}$ are constrained to be finite polygonal regions $\Delta_A$  and $\Delta_B$ inside ${\mathbb R}^s$ and ${\mathbb R}^{N-s}$ respectively, whose boundary are determined by the locus where the effective mass for the individual chiral multiplet vanishes, i. e. the root of Higgs branch. The dual photons then give ${\mathbb T}^{s}$ and ${\mathbb T}^{N-s}$ fibrations over $\Delta_A$ and $\Delta_B$, where the toroidal fibers can become singular at the boundaries of the polygons $\Delta_A$ and $\Delta_B$ when some or all of the $U(1)_J$ symmetries are restored there.  Together with the $\{\sigma_a\}$ and $\{\hat{\sigma}_p\}$, the Coulomb branches for Theory A and Theory B can both be identified as toric manifolds \cite{Dorey:1999rb}.
The evidence for the conjectured mirror symmetry between Theories A and B was then provided by realizing that the quotient construction for the classical Higgs branches also yield identical toric manifolds as the quantum corrected Coulomb branches \cite{Dorey:1999rb}. 
We can extend further the technique of gauging of the $U(1)_N \times U(1)_R$ R-symmetry to construct new $\cN=2$ non-Abelian mirror pairs from $\cN=4$ ones while however in this case the Coulomb branch receives further non-perturbative corrections in the superpotential due to three dimension instanton \cite{Tong:2000ky}, rendering the explicit checking of mirror symmetry by identifying the geometries of Coulomb and Higgs branches rather difficult.

\subsection{Interpolation to Two Dimensional Mirror Symmetry}
\paragraph{}
In two dimensions, there is also the well-known mirror symmetry for $\cN=(2,2)$ quantum field theories, which exchanges the vector $U(1)_V$ and axial $U(1)_A$ R-symmetries, and chiral and twisted chiral multiplets among the mirror pair. Explicitly in \cite{Hori:2000kt}, two dimensional mirror symmetry can be established through a generalization of scalar-scalar or T-duality \cite{Rocek:1991ps}. Starting with an Abelian gauged linear sigma model, one dualizes the phase of individual charged chiral multiplet into twisted multiplet, incorporates the dynamically generated superpotential due to two dimensional instantons or vortices, and shows that the mirror dual theory obtained is precisely a Landau-Ginzburg model with Toda-type superpotential given in terms of twisted multiplets.
This is rather similar to three dimensional $\cN=2$ Abelian gauge theories reviewed earlier, where the Coulomb branch moduli space metric which receives one-loop perturbative quantum corrections is mapped by three dimensional mirror symmetry to the Higgs branch of the dual theory, which is described by a sigma model with target space being the Higgs branch. 
It is thus natural to ask if the three and two dimensional mirror symmetries can be interpolated somehow through compactification. 
Naively this seems to be difficult as former is an infra-red duality, and additional quantum corrections may appear as theories flow to two dimensions. Indeed the subtleties for relating infra-red dualities in three and four dimensions have been extensively discussed in \cite{Aharony:2013dha, Aharony:2013kma}.   
However, given the recent progress on computing the exact partition functions of supersymmetric gauge theories on curved manifolds using localization techniques \cite{Kapustin:2009kz, Kapustin:2010xq, Hama:2010av, Hama:2011ea, Imamura:2011wg, Benvenuti:2011ga, Gomis:2012wy, Doroud:2012xw, Benini:2012ui}, not only we have been able to perform direct field theoretical precision tests of various infra-red dualities arising from string theory constructions, such as three dimensional mirror symmetry, but these exact results also enables us to explicitly check the interpolation of these dualities between different dimensions, as we will do next.

\section{Precision Checks of Mirror Symmetry in 2+1 and 1+1 Dimensions}\label{PrecisionCheck}
\subsection{Relating Partition Function on Ellipsoid $S_b^3$ to 3d Index on $S^2\times_q S^1$}\label{EllipCheck}
\paragraph{}
The partition functions of $\cN=2$ supersymmetric gauge theories computed on 3d ellipsoid $S_b^3$ have recently attracted significant amounts of interests. Starting with \cite{Hosomichi:2010vh} where the authors demonstrated that the building elements of the partition functions on round $S^3$ are related to Liouville correlatiors with $b=1$,  they further suggested that $b$ can parameterize the deformation of the background three sphere. This was realized in the subsequent work \cite{Hama:2011ea, Imamura:2011wg}, where the authors calculated the partition function on the $S^3_b$ by applying the localization techniques. We will readily apply these results in the precision test of the mirror symmetry between Theory A and Theory B reviewed earlier.

However, to demonstrate the interpolation between 2d and 3d mirror symmetries,
instead of the ellipsoid $S_b^3$, we need the partition function on $S^2\times_q S^1$ or the superconformal index, and then dimensionally reduce along $S^1$ fiber \footnote{Similar 3d and 2d dimension reduction was first considered in in \cite{Benini:2012ui} .}.
Fortunately, as studied extensively in \cite{Pasquetti:2011fj, Beem:2012mb}, there exists a more fundamental building element for the partition functions defined on different background geometries, called {\it "holomorphic block"}. 
We can build partition functions from these holomorphic blocks by applying certain {\it "gluing rules"}. 
Explicitly in \cite{Pasquetti:2011fj}, 
the author observed that in the squashing limits, i. e. taking the deformation parameter $b \to 0$, the ellipsoid $S^3_b$ degenerates into a twisted product of $R^2\times_q S^1$. At the leading order in $b$ expansion, the partition function can be interpreted as the equivariant partition function summing up the topological vortices and their zero mode fluctuations. The twisting or equivariant parameter $q$ here comes from the fugacity for the rotation in $R^2$ and is related to the deformation parameter of ellipsoid $S_b^3$ via $q=e^{2\pi i b^2}$.
In the complimentary limit $1/b\rightarrow 0$,  the resultant partition function on $S_b^3$ can also be interpreted as the equivariant partition function summing up the topological anti-vortices and their zero mode fluctuations\footnote{We will make precise the definition of ``vortex'' and ``anti-vortex'' by specifying the dependence of their action on the deformation parameter $b$.}. The author of \cite{Pasquetti:2011fj} then proposed that the ellipsoid partition function $Z_{S^3_b}$ can be factorized into
\be
Z_{S^3_b}(\{m, \zeta\})=\sum_{i=1} ||{\mathfrak B}^{i}(\{m,\zeta\})||^2_{\rm S}
\ee
where the symbolic summation here should be understood as summing over different possible convergent contours or vacua,
and the subscript $S$ here indicates the way two blocks, $\fB^{i}(\{m,\zeta\})$ and its anti-holomorphic partner, are fused together which will be specified momentarily. 

For finite $b$, it was further shown in \cite{Pasquetti:2011fj} and explained in \cite{Beem:2012mb} that the factorization property of partition functions still holds.
Indeed both $S^2\times_q S^1$ and $S^3_b$ geometries can be regarded as a two-torus $T^2$ non-trivially fibered over an interval, such that the cycles of $T^2$ degenerate at the end of the interval while the interval is stretched to infinite.
At the two ends of the interval, partition functions can now be decomposed into two copies of the partition function on $D^2\times_q S^1$ \cite{Beem:2012mb}, where $D^2$ is a semi-infinite cigar fibered over $S^1$ with $U(1)$ holonomy $q$, and the resulting geometry is the well-known ``Melvin's cigar''. In the large radius limit for $D^2$, we can recover the vortex/anti-vortex partition function on $R^2\times_q S^1$.
Two copies of partition function on $D^2\times_q S^1$ can now be fused together using different elements of the modular group $SL(2, {\mathbb Z})$ acting on the complex structure $\tau$ of $T^2$ followed by reversing the relative orientation.   
Respectively the identity element $\tau \to -\tau$ hence called `` {\it id}-fusion '', yields the partition function on $S^2\times_q S^1$; while the S-element $\tau \to \frac{1}{\tau}$, hence called ``S-fusion'', yields the partition function on $S_b^3$, for which the negative sign here accounts for orientation reversal.
Concluding from these, we can start from the partition on $S_{b}^3$ and explicitly factorize it into holomorphic blocks for a given gauge theory. We then apply the identity fusion to obtain the corresponding partition function on twisted product $S^2\times_q S^1$ and perform the desired dimensional reduction \footnote{Factorization of 3d superconformal index was also considered in \cite{Hwang:2012jh}}.

\subsection{Checking Mirror Symmetry on Ellipsoid $S_b^3$}
\paragraph{}
Following the results in \cite{Hama:2011ea}\cite{Imamura:2011wg}, we can readily write down the partition functions on $S_b^3$ for the mirror pair Theory A and Theory B specified  in section \ref{3dMirrorSym}. 
The precision test for the mirror symmetry can be consequently performed by directly matching their partition functions. 

Let us first write down the partition functions for Theory A and Theory B defined on ellipsoid using the results in \cite{Hama:2011ea}:
\begin{equation}\label{PartThA}
Z^{\rm A}_{S^3_b}(m_i; \zeta^a)=\int 
 d^s\sigma 
 \prod^N_{i=1}\left(s_b\left(c_b-\sum^s_{a=1}Q^a_i\sigma_a-m_i\right)\right)
 \exp\left(-2\pi i \sum_{a=1}^{s}\zeta^a\sigma_a \right)
 \exp{\left(\frac{-i\pi}{2}\sum\limits^N_{i=1}
 \sum\limits^s_{a=1}
 \sum\limits^s_{b=1}Q_i^aQ_i^b\sigma_a\sigma_b\right)},
\end{equation}
\begin{equation}
\begin{split}\label{PartThB}
Z^{\rm B}_{S_b^3}(\hat{m}_i; \hat{\zeta}^p)
=\int &
 d^{N-s}\hat{\sigma} 
 \prod^N_{i=1}\left(s_b\left(c_b-\sum^{N-s}_{p=1}\Q^p_i\hat{\sigma}_p-\m_i\right)\right)
 \exp{\left(-2\pi i\sum_{p=1}^{N-s} \z^p\hat{\sigma}_p \right)}
 \exp{\left(\frac{i\pi}{2}\sum\limits^N_{i=1}
 \sum\limits^{N-s}_{p=1}
 \sum\limits^{N-s}_{q=1}
 \Q_i^p\Q_i^q\hat{\sigma}_p\hat{\sigma}_q\right)},
\end{split}
\end{equation}
where the "double-sine" function $s_b(x)$ which will play important role in our discussion is defined as:
\begin{eqnarray}
\label{doublesine}
&&s_b(x)=\prod^{\infty}_{m,n=0}\frac{mb+n/b-i c_b-ix}
{mb+n/b -i c_b+ix}
=\frac{e^{-i\frac{\pi}{2}x^2}}{\prod_{r=0}^{\infty}\left(1+e^{2\pi b x} q_1^{-(2r+1)}\right) \left(1+e^{\frac{2\pi}{b}x}q_2^{-(2r+1)}\right)}\nn\\
&& c_b=\frac{i}{2}\left(b+\frac{1}{b}\right), \quad q_1=e^{i\pi b^2}, \quad q_2=e^{i\frac{\pi}{b^2}}.
\end{eqnarray}
Let us comment on various notations entering \eqref{PartThA} and \eqref{PartThB}. 
Here $\sigma_a,~ a=1,\dots, s$ and $\hat{\sigma}_p,~ p=1,\dots N-s, $ are respectively the real scalar components in the $U(1)_a \subset U(1)^s$ and $U(1)_p\subset U(1)^{N-s}$ vector multiplets.
The first terms in the integrand of \eqref{PartThA} come from $N$ chiral multiplets, and the middle terms are from FI terms denoted by $\{\zeta^a\}$ while the last terms are from Chern-Simons terms, likewise for the various terms in integrand in \eqref{PartThB}.

To demonstrate the three dimensional mirror symmetry, we would like to match their partition functions \eqref{PartThA} and \eqref{PartThB}. This is seeming difficult as they are integrals involving different number of variables. However, the following identity for the Fourier transform for $s_b(x)$ becomes useful:
\be
\label{Fourier}
\int dz\, e^{\frac{i\pi}{2}z^2}e^{i\pi c_bz}e^{2\pi i wz}s_b(c_b+z).
=
s_b(c_b+w)e^{\frac{-i\pi}{2}w^2}e^{-i\pi c_b w}.
\ee
Now let us consider ellipsoid partition function for Theory A \eqref{PartThA} and pick out $s$ of $N$ massive chiral multiplets, for which we can package their charges into a $s\times s$ matrix $(Q_u)^a_b$ and their real masses as a $s$-column vector $m_u$; while for the remaining $(N-s)$ chiral multiplets we denote their charges as a $(N-s)\times s$ matrix $(Q_d)^p_a$ and their real masses as a $(N-s)$ column vector. We can define now $s$ new integration variables: 
\be
y_a = -(Q_u \sigma+m_u)_a, \quad a =1,2, \cdots s  
\ee
and also use \eqref{Fourier} to rewrite the double sine functions for the remaining $N-s$ chiral multiplet contributions into integrals, obtaining:
\begin{eqnarray}
\label{integrand ThA}
&&Z^{\rm A}_{S_b^3}(m_i;\zeta^a)=
\sum_{\{Q_u\}}
\frac{e^{i\pi\left( 
-c_b{I^{\rm T}_{N-s} m_d} + 
\frac{{m_d^Tm_d}}{2}   
+2\zeta^{\rm T} Q^{-1}_um_u
+({I^T_{N-s}}c_b-{m_d^T}){Q_dQ_u^{-1}m_u}  -\frac{{m_u^Tm_u}}{2}    \right)}}{|\det(-Q_u)| }
\int d^{N-s}x_pd^sy_a
\prod^s_{a=1}s_b(c_b+y_a)
\prod^N_{p=s+1}s_b(c_b+x_p)
\nn\\
&&\times
e^{i\pi\left( 
{\sum\limits^s_{a=1} y^a\left(2{\zeta^T (Q_u^{-1})}     
+(c_b{I^T_{N-s}}-{m_d^T}){ (Q_dQ_u^{-1}) }-{m_u^T} \right)_a               }
+{\sum\limits^{N}_{p=s+1}x^p\left(c_b{I^T_{N-s}}-2{m_d^T}+2{(Q_d Q_u^{-1}m_u)^T}\right)_p}\right)
 -
 i\pi\left( \sum\limits^s_{a=1} \frac{y_a^2}{2} 
-\sum\limits^N_{p=s+1}\frac{x_p^2}{2}
-2\sum\limits^s_{a=1} 
  \sum\limits^N_{p=s+1} 
  x_p(Q_dQ_u^{-1})_{a}^p y^a
  \right) .
 }
\nn\\
\end{eqnarray}
We can perform similar rewriting for the ellipsoid partition function of Theory B \eqref{PartThB}. Choosing $(N-s)$ out of $N$ chiral multiplets and packaging their charges into $(N-s)\times (N-s)$ matrix $(\Q_d)^r_p$, their real masses can be packaged as a $(N-s)$-column vector $\m_d$, we can define now $N-s$ new integration variables:
\be
z_p=-(\Q_d \hat{\sigma}+\m_d)_p.
\quad p=1,2,\cdots N-s
\ee 
While the charges of remaining chiral multiplets can be packaged into $(N-s)\times s$ matrix $\Q_u$ and their real masses $s$-column vector $\m_u$. The resultant integral expression using \eqref{Fourier} becomes:
\begin{eqnarray}
\label{integrand ThB}
&&Z^{\rm B}_{s_b^3}=\sum_{\{\Q_d\}}
\frac{e^{i\pi\left( 
 -\frac{{\hat{m}_u^T\hat{m}_u}}{2} 
 +c_b{I^T_{k}\cdot \hat{m}_u} 
 -({I^T_{k}}c_b-{\hat{m}_u^T}){(\hat{Q}_u\hat{Q}_d^{-1})\hat{m}_d} 
 +\frac{{\hat{m}_d^T\hat{m}_d}}{2}    
 +2\hat{\zeta}^T\cdot (\hat{Q}_d)^{-1}\hat{m}_d 
\right)}}{|\det(-\Q_d)|}
\int{d^{s}w_ad^{N-s}z_p}
\prod^s_{a=1}s_b(c_b+w_a)\prod^N_{p=s+1}s_b(c_b+z_p)
\nn\\
&&\times e^{i\pi \left(
\sum\limits^N_{p=s+1} z^p 
(2 \hat{\zeta}^T \hat{Q}_d^{-1}
+(-c_bI^T_s+ \hat{m}_u^T )\cdot(\hat{Q}_u\hat{Q}_d^{-1}))_p
+
\sum \limits_{ a=1 }^s
 w^a (-c_bI^T_s+2\hat{m}_u^T-2\hat{m}_d^T (\hat{Q}_d^{-1})^T\hat{Q}_u^T)_a\right)
- i\pi\left( \sum\limits^s_{a=1} \frac{w_a^2}{2} 
-\sum\limits^N_{p=s+1}\frac{z_p^2}{2}
+2\sum\limits^s_{a=1} 
  \sum\limits^N_{p=s+1} 
  w_a(\hat{Q}_u\hat{Q}_d^{-1})_{p}^a z^p
  \right) 
 }\nn\\
\end{eqnarray}
The summations over different $\{Q_u\}$ and $\{\Q_d\}$ in \eqref{integrand ThA} and \eqref{integrand ThB} indicate respectively that we are summing over all possible choices of picking out $s$ and $N-s$ out of $N$ chiral fields, yielding equal number of terms.
This is equivalent to summing over different possible discrete vacua in Theory A and Theory B. Happily they are of equal number as required for three dimensional mirror symmetry.
Let us now focus on the $N$ variable integrals in \eqref{integrand ThA} and \eqref{integrand ThB}. We can first impose the charge orthogonality condition \eqref{charge relation}  where now we specify on a pair of vacua so that it can be written as $(Q_u^T\Q_u +Q_d^T \Q_d)^a_p=0$. Also, we impose the following generalization of the proposed map between the parameters \eqref{MirrorMap} on curved manifolds:
\begin{eqnarray}
\label{mirror comparing 1}
&&(Q_u^{-1} \zeta_{\rm eff})^a= (\m_u-(\Q_u\Q_d^{-1})\m_d)^a, \\
\label{mirror comparing 2}
&&(\Q_d^{-1} \hat{\zeta}_{\rm eff})^p=-(m_d-(Q_dQ_u^{-1})m_u)^p.
\end{eqnarray}
Here we have generalized the effective FI parameters to $\{\zeta_{\rm eff}^a; \hat{\zeta}_{\rm eff}^p\}$:
\begin{eqnarray}\label{cFIeffA}
&&\zeta_{\rm eff}^a=\zeta^a-\frac{1}{2}\sum_{i=1}^N Q_{i}^a (m^i-c_b), \\ 
\label{cFIeffB}
&&\hat{\zeta}_{\rm eff}^p=\hat{\zeta}^p+\frac{1}{2}\sum_{i=1}^N\Q^p_i(\m_i-c_b),
\end{eqnarray}
where the $c_b$ dependent terms above come from the Chern-Simons coupling of gauge fields with the background R symmetry. 
We can eliminate such terms by imposing $\sum_{i=1}^N Q_i^a=0$ and $\sum_{i=1}^N \Q_i^p=0$, which are precisely the Calabi-Yau conditions for Theory A and Theory B. 
With these constraints and mapping of parameters, after identifying the integration variables as $\{x_p\} \leftrightarrow \{z_p\}$ and $\{y_a\} \leftrightarrow \{w_a\}$, we can now show that  the two integrals in \eqref{integrand ThA} and \eqref{integrand ThB} can be precisely identified!

Now consider the overall constant phase factors in \eqref{integrand ThA} and \eqref{integrand ThB} which are generated from the chiral fermions charged under background flavor and R-symmetries\footnote{We will rewrite them in more apparent form momentarily.}. 
We can cancel these anomaly terms by including the corresponding background Chern-Simons couplings \cite{Beem:2012mb}.
After doing so, we can match the remaining phase terms in \eqref{integrand ThA} and \eqref{integrand ThB} which can be packaged as ${2\pi i (\zeta_{\rm eff}Q_u^{-1} m_u)}$ and ${2\pi i (\hat{\zeta}_{\rm eff}\Q_d^{-1} \hat{m}_d)}$ by cross multiplying \eqref{mirror comparing 1} and \eqref{mirror comparing 2} respectively with $m_{u,a}$ and $\m_{d, p}$ and imposing the orthogonality condition $\sum_{i=1}^N m_i \hat{m}^i = 0 $. Finally, for the overall constant factors arising from the Jacobians $1/|\det (-Q_u)|$ and $1/|\det(-\hat{Q}_d)|$, we show in Appendix \ref{determinant} that the ratio of these terms between two mirror theories remain the same among different choices of charge $\{Q_u\}$ and $\{\hat{Q}_d\}$, as long as the orthogonal condition \eqref{charge relation} can be rewritten as  $(Q_u^T\Q_u +Q_d^T \Q_d)^a_p=0$, for which we then scale the charges to set this ratio equals to 1.
This establishes the matching of the $S_b^3$ partition functions of Theory A and Theory B and the non-trivial verification of three dimensional mirror symmetry.

To explicitly evaluate the integrals in \eqref{integrand ThA} and \eqref{integrand ThB}, we promote $\{x^p, y^a; w^a, z^p\}$ to complex variables and close the integration contour to the upper half complex plane to pick out the residues associated with the simple poles in $s_b(x)$. For \eqref{integrand ThA}, they are respectively given by:
\be
y_a = i\left(f_a b+\frac{g_a}{b}\right), \quad  x_p = i\left(\alpha_p b+\frac{\beta_p}{b}\right),\quad f_a, g_a, \alpha_p, \beta_p = 0, 1, 2, \dots,
\ee
and identical pole conditions are given for $\{w_a, z_p\}$ integrations in \eqref{integrand ThB}.
The final result becomes: 
\begin{eqnarray}
\label{Theory A}
&&Z^{\rm A}_{S^3_b}(m_i ; \zeta^a)
=
e^{ \frac{i\pi}{2}\sum_{i=1}^N (m_i+c_b)^2 +\frac{i\pi}{2}N c_b^2}
\sum_{\{Q_u\}}
\frac{(2\pi)^N}{\left|\det(-Q_u)\right|}
\prod_{a=1}^s e^{2\pi i  (\zeta_{\rm eff} Q_u^{-1})_a m_u^a}
\nn\\
&&\times \left[
\sum _{ \{ { \alpha  }_p,\beta_p, f_a, g_a=0 \} }^{ \infty  } 
\prod_{a=1}^s \prod_{p=s+1}^N
{ \frac { 
e^{ 
-2\pi (\zeta_{\rm eff} Q_u^{-1})_a (f^a b+g^a /b)  
} }
{ { \prod \limits_{ \delta =1 }^{ { f }_{ a } }{ (1-{ q }_{ 1 }^{ -2\delta  }) }  } 
{ \prod\limits _{ \Delta =1 }^{ { g }_{ a } }{ (1-{ q }_{ 2 }^{ -2\Delta  }) }   }  }  } 
 \frac { 
 { e }^{ -2\pi { { ((Q_d Q_u^{-1})m_u-m_d ) }_{ p}(\alpha^p b +\beta^p/b) }  } }
  {  \prod \limits_{ m=1 }^{ { \alpha  }_{ p } }{ (1-{ q }_{ 1 }^{ 2m })\prod \limits_{ n=1 }^{ { \beta  }_{ p } }{ (1-{ q }_{ 2 }^{ 2n })}   }  
 }
 e^{ 
-2\pi i(\alpha_p b+\beta_p/b)(Q_d Q_u^{-1})^{p}_a(f^a b+g^a/b)
 }\right] . 
 \nn\\
 &&=
e^{ \frac{i\pi}{2}\sum_{i=1}^N (m_i+c_b)^2 +\frac{i\pi}{2}N c_b^2}
\sum_{\{Q_u\}} \sum _{ \{ f_a, g_a=0 \} }^{ \infty  } 
\frac{(2\pi)^N}{\left|\det(-Q_u)\right|}
\prod_{a=1}^s \prod_{p=s+1}^N e^{2\pi i  (\zeta_{\rm eff} Q_u^{-1})_a m_u^a}
\nn\\
&&\times \left[
{ \frac { 
e^{ 
-2\pi (\zeta_{\rm eff} Q_u^{-1})_a (f^a b+g^a /b)  
} }
{ { \prod \limits_{ \delta =1 }^{ { f }_{ a } }{ (1-{ q }_{ 1 }^{ -2\delta  }) }  } 
{ \prod\limits _{ \Delta =1 }^{ { g }_{ a } }{ (1-{ q }_{ 2 }^{ -2\Delta  }) }   }  }  } 
\frac{1}{\left(e^{-2\pi b[((Q_dQ_u^{-1})m_u-m_d)_p-i(Q_dQ_u^{-1})^a_p(f_a b+\frac{g_a}{b})]}; q_1^2\right)_{\infty}
\left(e^{-\frac{2\pi}{b}[((Q_dQ_u^{-1})m_u-m_d)_p-i(Q_dQ_u^{-1})^a_p(f_a b+\frac{g_a}{b})]}; q_2^2\right)_{\infty}
}
\right] . 
 \nn\\
 \end{eqnarray}
\bea
\label{Theory B}
&&Z^{\rm B}_{S^3_b}(\hat{m}_i; \hat{\zeta}^p)
=e^{-i\frac{\pi}{2}\sum_{i=1}^N(\m_i-c_b)^2-\frac{i\pi}{2}Nc_b^2}\sum_{\{\Q_d\}}\frac{(2\pi)^N}{|\det(-\hat{Q}_d)|} 
\prod_{p=s+1}^{N} e^{2\pi i (\hat{\zeta}_{\rm eff}\Q_d^{-1})_p \m^p_d}
\nn\\ 
&&\times \left[
\sum _{ \left\{\alpha_p, \beta_p ,f_a, g_a=0 \right\} }^{ \infty  }
\prod_{a=1}^s\prod_{p=s+1}^N
{ \frac { 
e^{ 
-2\pi 
{ (\hat{\zeta}_{\rm eff} \Q_d^{-1})_p(\alpha^p b+\beta^p/b)
}  } }
{ { \prod \limits_{ \delta =1 }^{ \alpha_p }{ (1-{ q }_{ 1 }^{ 2\delta  }) }  } 
{  \prod\limits _{ \Delta =1 }^{ \beta_p }{ (1-{ q }_{ 2 }^{ 2\Delta  }) }  }  }  } 
 \frac { 
 e^{ -2\pi {{ (\m_{ u }-(\Q_u\Q_d^{-1})\m_d )}_a  (f^a b+g^a/b)  }  } }
 { {  \prod \limits_{ m=1 }^{ f_a }{ (1-{ q }_{ 1 }^{ -2m })
       \prod \limits_{ n=1 }^{ g_a}{ (1-{ q }_{ 2 }^{ -2n }) } }   }  }  
 e^{ 2\pi i (f_a b + g_a/b)(\Q_u \Q_d^{-1})^a_p (\alpha^p b + \beta^p/b) }
 \right]  
\nn\\
&&=e^{-i\frac{\pi}{2}\sum_{i=1}^N(\m_i-c_b)^2-\frac{i\pi}{2}Nc_b^2}\sum_{\{\Q_d\}}
\sum _{ \left\{f_a, g_a =0 \right\} }^{ \infty  }\frac{(2\pi)^N}{|\det(-\hat{Q}_d)|} 
\prod_{p=s+1}^{N} e^{2\pi i (\hat{\zeta}_{\rm eff}\Q_d^{-1})_p \m^p_d}
\nn\\ 
&&\times \left[ 
 \frac { 
 e^{ -2\pi {{ (\m_{ u }-(\Q_u\Q_d^{-1})\m_d )}_a  (f^a b+g^a/b)  }  } }
 { {  \prod \limits_{ m=1 }^{ f_a }{ (1-{ q }_{ 1 }^{ -2m })
      \prod \limits_{ n=1 }^{ g_a}{ (1-{ q }_{ 2 }^{ -2n }) } }   }  }  
\frac{1}{\left(e^{-2\pi b(\hat{\zeta}_{\rm eff}\hat{Q}_d^{-1})_p+2\pi i(\Q_u\Q_d^{-1})^a_p(f_a b+\frac{g_a}{b})}; q_1^2\right)_{\infty}
\left(e^{-\frac{2\pi}{b}(\hat{\zeta}_{\rm eff}\hat{Q}_d^{-1})_p+2\pi i(\Q_u\Q_d^{-1})^a_p(f_a b+\frac{g_a}{b})}; q_2^2\right)_{\infty}
}
 \right].
 \nn\\  
\eea
Here we have used the effective FI, $\{\zeta_{\rm eff}^{a}; \hat{\zeta}_{\rm eff}^p\}$, defined in \eqref{cFIeffA} and \eqref{cFIeffB} to rewrite the constant phases in \eqref{integrand ThA} and \eqref{integrand ThB} and express the resultant summations.
In the second equality in each expression above, we have used the identity for Pochhammer symbol $\sum_{n=0}^{\infty}\frac{x^n}{(q;q)_{\infty}}=\frac{1}{(x;q)_{\infty}}$ \cite{Gasper} to replace the summation over $\{\alpha_p, \beta_p\}$. Such rewriting is needed for later discussions on factorization of ellipsoid partition functions and comparison with calculation of equivariant vortex partitions.

Having matched the integral representation of the partition functions for mirror pairs in \eqref{integrand ThA} and \eqref{integrand ThB},
let us briefly mention another perspective to view the three dimensional mirror symmetry following \cite{Kapustin:1999ha}.
Due to the presence of $e^{-2\pi i\zeta^a\sigma_a}$ and $e^{-2\pi i \hat{\zeta^p}\hat{\sigma}_p}$, another way to view the partition functions \eqref{PartThA} and \eqref{PartThB} is to regard them as Fourier transforms of the remaining integrands denoted as $\cF^{\rm A}(\sigma_a; m_i)$ and $\cF^{\rm B}(\hat{\sigma}_p; \hat{m}_i)$ for the time being, with the $\{ \zeta^a\}$ and $\{\hat{\zeta}^p\}$ being the dual variables.
Starting from \eqref{PartThA},  we can now perform the Fourier inverse transform with respect to $s$ FI parameters $\{\zeta^a\}$ to obtain $\cF^{\rm A}(\sigma^a; m_i)$. 3d mirror symmetry however implies the equivalence of $Z^{\rm A}(\zeta^a; m_i)$ and $Z^{\rm B}(\hat{\zeta^p}; \hat{m}_i)$, which allows us to deduce the following relation between $\cF^{\rm A}(\sigma^a; m_i)$ and $\cF^{\rm B}(\hat{\sigma}^p; \hat{m}_i)$:
\be
\label{Fourier Tansform}
\cF^{\rm A}(\sigma_a; m_i)=\int d^s \zeta \int d^{N-s}\hat{\sigma} \exp\left[2\pi i\left( \sum_{a=1}^s \zeta^a\sigma_a-\sum_{p=1}^{N-s}\hat{\zeta}^p\hat{\sigma}_p \right)\right]
\cF^{\rm B}(\hat{\sigma}_p, \hat{m}_i).
\ee
One should however note that in performing the integration, we should remember $\{\zeta^{a}\}$ should be treated as a function of the real mass parameters $\{\hat{m}_i\}$ as given by the mirror map \eqref{MirrorMap}, or more precisely picking out $N-s$ linear combinations out of $\{\hat{m}_i\}$.  Moreover, reviewing  \eqref{integrand ThA} and \eqref{integrand ThB}, rather than merely a mathematical intermediate step, applying Fourier Transformation holds a more physical interpretation. From \cite{Kapustin:1999ha}, what we did above is the "piecewise Fourier Transformation", introducing additional integration variables $\{x_p\}$ and $\{w_a\}$ which correspond to three dimensional generalization of the "twisted vector multiplets". As a result, the expressions \eqref{integrand ThA} and \eqref{integrand ThB} can be identified with the same BF-theory, where the BF-terms $(Q_dQ_u^{-1})^p_a$ and $(\Q_u\Q_d^{-1})^a_p$ couple the vector multiplets with the twisted vector multiplets. 

As discussed in the section \ref{3dMirrorSym}, we can also obtain Theory A and Theory B from  $\cN =4$ mirror pair by gauging the diagonal $U(1)_{R-N}$
and integrating out the heavy fermions which acquire large real masses. We briefly demonstrate that such procedures can also be performed within ellipsoid partition functions.
For example, starting with $\cN=4$ Theory A', we can also write down the $S_b^3$ partition function:
\begin{equation}
Z^{\rm A'}_{S^3_b}
=\prod\limits^s_{a=1}s_b(-\mu +c_b) 
\int d^s\sigma \prod\limits^N_{i=1}
\left(
\frac{
{s_b\left(c_b-\sum\limits^s_{a=1}  Q^a_i\sigma_a -m_i'-\mu  \right)}}
{{s_b\left(-c_b-\sum\limits^s_{a=1}  Q^a_i\sigma_a -m_i'+\mu \right)}}
\right)
e^{-2i\pi \sum_{a=1}^{s} \zeta^a\sigma_a}.
\end{equation}
Here the $\{s_b(-\mu+c_b)\}$ come from the neutral chiral multiplets making up $\cN=4$ vector multiplets, and the $\{s_b(-c_b-\sum\limits^s_{a=1}  Q^a_i\sigma_a -m'_i+\mu)\}$ in the denominator account for the additional $\cN=2$ anti-fundamental chiral multiplets making up $\cN=4$ hypermultiplets. 
Let us now define the shifted mass $m_i=m_i'+\mu$ and take the limit $\mu\to \infty$ while keeping fixed $m_i$. 
In such limit, $\cN=4$ supersymmetry is partially broken so that we can decouple the heavy neutral and anti-fundamental chiral multiplets in the partition function, and their only remaining contributions can be deduced from the asymptotic behavior of the double-sine function:
$$
\lim _{x\to -\infty} s_b(c_b+x) \rightarrow  e^{-\frac{i\pi}{2}(c_b+x)^2}.
$$  
As a result, the neutral chiral multiplets merely contribute to an overall constant while the anti-fundamental chiral multiplets contribute Chern-Simons terms:
\begin{equation}
\begin{split}
\lim_{\mu\rightarrow \infty} &Z^{\rm A'}_{S^3_b}
\\
\sim
&\int d^s\sigma \prod\limits^N_{i=1}
\left(
{s_b\left(c_b-\sum\limits^s_{a=1}  Q^a_i\sigma_a -m_i \right)}
\right)
e^{\frac{-i\pi}{2}
\sum\limits^{N}_{i=1}\left(\left(2(c_b+m_i)-4\mu\right)
\sum\limits^s_{a=1}Q^a_i\sigma^a\right)}
e^{-2i\pi \zeta^a\sigma^a}
e^{\frac{-i\pi}{2}\left(\sum\limits^s_{a=1}Q^a_i\sigma^{a '}\right)^2}
\\
=
&\int d^s\sigma \prod^N_{i=1}
\left(s_b\left(c_b-\sum^s_{a=1}Q^a_i\sigma_a-m_i\right)\right)
 \exp\left(-2\pi i \sum_{a=1}^s\zeta^a\sigma_a \right)
 \exp{\left(\frac{-i\pi}{2}\sum\limits^N_{i=1}
 \sum\limits^s_{a=1}
 \sum\limits^s_{b=1}Q_i^aQ_i^b\sigma_a\sigma_b\right)}
\end{split}
\end{equation}
where we have also rescaled the FI parameter $\zeta^a=\zeta^{a '} +\sum\limits^N_{i=1}(\frac{1}{2}(c_b+m_i)-\mu)Q^a_i$. We precisely recover the $\mathcal{N}=2$ Theory A in \cite{Aganagic:2001uw} and the same partition function given in \eqref{PartThA}. We can also repeat identical steps to the $S_b^3$ function for Theory B' to obtain the corresponding partition function for Theory B \eqref{PartThB}. Therefore, if starting from the matching between the $S_b^3$ partition functions for an $\cN=4$ mirror pair, we can equivalently apply the procedure of gauging R-symmetry to derive the matching of the correspondingly $\cN=2$ mirror pair.

\subsection{Factorization, Identity Fusion and Reduction to Two Dimensions}
\paragraph{}
In this section, we will first factorize the ellipsoid partition functions for our mirror pair computed in the previous section  into so-called {\it holomorphic blocks} and apply the identity fusion method to obtain the partition functions on  $S^2\times_q S^1$ or the superconformal index \cite{Beem:2012mb}. 
By reducing along $S^1$ to two dimensions, we then demonstrate the interpolation between mirror symmetries in three and two dimensions \cite{Aganagic:2001uw}.
To provide further evidence for such interpolation,  we also explicitly rewrite the $S^2\times_q S^1$ partition function in terms of q-deformed generalization of Bessel functions which in two dimension limit reduces to the $S^2$ partition function of the corresponding $\cN=(2,2)$ Landau-Ginzburg model computed in \cite{Gomis:2012wy}, which is the two dimensional mirror dual of  gauged linear sigma model.

\subsubsection{Holomorphic Block and Changing to $S^2 \times_q S^1$}
\paragraph{}
Here we will focus primarily on Theory A, as the same steps and arguments can be equivalently applied to Theory B. 
Let us begin by demonstrating that via factorization the ellipsoid partition function $Z_{S^3_b}^{\rm A}(m_i, \zeta^a)$ can be split 
into product of purely $b$ dependent terms and purely $1/b$ dependent terms, i. e. ``holomorphic block" and  ``anti-holomorphic block'', and they are related via S-transformation  mapping $b\to 1/b$.  
It can be seen from the second equality in \eqref{Theory A} that $Z_{S_b^3}^{\rm A}(m_i,\zeta^a)$ is almost manifestly of factorizable form, except the phase factors $\exp({2\pi i (Q_d Q_u^{-1})^a_p f_a})$ and $\exp({2\pi i(\Q_u \Q_d^{-1})^a_p g_a})$ in the denominator,
which need to be discussed.  
Clearly, for generic values of $f_a, g_a$,  the factors $\exp({2\pi i(Q_d Q^{-1}_u)^{a}_p f_a})$ and $\exp({2\pi i(Q_d Q^{-1}_u)^{a }_p g_a})$ yield different phases and render the factorization into purely $b$ and $1/b$ dependent pieces difficult. 

Let us consider the simplest case of $s=1$ and all the $Q_i$ are mutually prime so that chosen $Q_u$ is a number and $(Q_d)^p$ becomes a column vector. We can divide the set of non-negative integers $\{f\}$ into $|Q_u|$ sets of non-negative integers such that: 
\be\label{fracNumber}
\{f/Q_u\}=\{\hat{f}\} ; \{\hat{f} +1/Q_u\}; \{\hat{f}+2/Q_u\}, \dots ; \{\hat{f}+(Q_u-1)/Q_u\}, ~ \hat{f}=0, 1,2,3 \dots , 
\ee
under which now the factor $\exp({2\pi i {(Q_d)_p}f/{Q_u} })$ yields {\it identical} value for each set of integers. Dividing $\{g\}$ in identical way, we can now readily factorize the expression in \eqref{Theory A} into sum of $|Q_u|$ distinct products of holomorphic and anti-holomorphic blocks with exclusive $b$ and $1/b$ dependence, each summing over the different sets of non-negative integers. 
Let us recall holomorphic block can also be regarded as exponentiation of the on-shell value for the one loop exact twisted superpotential on $R^2\times_q S^1$ plus additional $S^1$ radius $R$ dependent corrections \cite{Pasquetti:2011fj, Beem:2012mb}. 
Using the results computed in \cite{Witten:1993yc, Hori:2000kt} for strictly $R\to 0$ limit, charge $Q_u$ chiral yields $|Q_u|$ equivalent vacua given by the minimization equation of the form $(Q_u \sigma + m_u)^{Q_u} = \exp(2\pi i\tau) $\footnote{We thank Toshiaki Fujimori for discussing this point.}. Each vacuum admits the fractional vortices whose vortex numbers are precisely given by different sets in \eqref{fracNumber}. For finite $R$, we simply include KK-modes and other $R$-dependent corrections. However, this does not change the allowed fractional vortex numbers in each vacuum. When constructing the ellipsoid partition function for a given $Q_u$, we are precisely summing over the $S$-fused product of the exponential of the holomorphic and anti-holomorphic twisted superpotentials, evaluated at each of these $|Q_u|$ vacua with different set of vortex numbers.  

Now we returning to general case of $U(1)^s$. Assuming that for each $U(1)_a$ the entries of $Q_i^a$ are mutually prime,
we can similarly divide the mode number $\{f_a\}$ and $\{g_a\}$ into following distinct sets given by:            
\bea\label{FactorPole}
&& \{ (Q_u^{-1})^c_a f_c\} \equiv \{ \hf_a + (Q_u^{-1})^c_a k_c\}; \quad \hf_a=0,1,2,3, \dots   \\
&&  \{ (Q_u^{-1})^c_a g_c\} \equiv \{ \hg_a + (Q_u^{-1})^c_a l_c\}; \quad \hg_a=0,1,2,3, \dots,
\eea
where, for fixed indices $a$ and $c$, the values of $k_c, l_c$ are chosen to be $k_c, l_c = 0, 1, 2, \dots (Q_u)^a_c - 1$.
For each set of $\{f_a\}$ and $\{g_a\}$, we can now factorize the ellipsoid partition function $Z_{S^3_b}(m_i; \zeta^a)$ into exclusively $b$ and $1/b$ dependent parts, exchanged under $b\leftrightarrow 1/b$ transformation.
Let us first define the following new variables: 
\begin{equation}
\label{variable}
\begin{split}
q=e^\hbar=e^{2\pi ib^2}=q^2_1, \quad
x_a=e^{2\pi b m_{u,a}},
\quad
x_p=e^{2\pi b m_{d,p}},\quad
y_a= e^{2\pi b\zeta_{{\rm eff}, a} },\quad
\by_a= e^{2\pi b(\zeta_{\rm eff} Q_u^{-1})_a}. 
\end{split}
\end{equation}
For a given $Q_u$ and specific vacuum with  $\{f_a; g_a\} = \{(Q_u)^c_a \hf_c; (Q_u)^c_a \hg_c \}$, the holomorphic block can then readily be expressed as \footnote{We will also perform the summation over $\{\alpha_p\}$ momentarily when we consider the interpretation of holomorphic block as K-theory vortex partition function.
Here we can also insert the background Chern-Simons coupling discussed before to cancel the first term coming from the pre-factor in \eqref{Theory A}.} :
\begin{eqnarray}\label{HBlock1}
&&\fB^{\rm A}{\left[\{x_i\},\{\ \by_a\}, Q_u;q\right]}\nn\\
&&=
\left[
\prod\limits^N_{i=1}
\theta ((q^{-1}x_i)^{\frac{1}{\sqrt{2}}};q)
\right]
\left[\prod\limits_{a=1}^{s}
\frac{\theta (\by_a x_a;q)}{\theta (\by_a;q)\theta (x_a;q)}\right]\nn
\left[
\sum\limits^\infty_{\{\alpha_p=0\}}
\sum\limits^\infty_{\{\hf_a=0\}}
\prod_{a=1}^{s}\prod_{p=s+1}^N
\frac{ 
y_a^{-\hf^a}
(x_p x_a^{-(Q_d Q_u^{-1})^a_p})^{\alpha^p}
q^{ -\alpha_p
 (Q_d)_a^p \hf^a}
}
{{  \prod \limits_{ \delta =1 }^{  (Q_u)^c_a \hf_c }{ (1-q^{ -\delta  }) }  }
{ \prod \limits_{ m=1 }^{ \alpha_p} (1-q^{ m })   }  
}\right],\nn\\
\end{eqnarray}
where we have replaced the summation index $\{f_a\}$ by $\{\hf_a\}$. In the above $\{x_i\}\equiv \{x_a; x_p\}$, and the theta-function is defined as
\be\label{theta function}
\theta(z;q)\equiv (-q^{\frac{1}{2}} z;q )_\infty (-q^{\frac{1}{2}} z^{-1};q)_\infty .
\ee
In obtaining \eqref{HBlock1}, we have also used the property of theta function \cite{Beem:2012mb} :
\begin{equation}
i^\#C^\# \exp \left[
-\frac{1}{2\hbar}\left(
(a\cdot X)^2+(i\pi +\frac{\hbar}{2})b(a\cdot X))
\right)
\right]
=
\left|\left|\theta ((-q^{\frac{1}{2}})^bx^a;q)\right|\right|^2_S,\quad x=\exp(X).
\end{equation}
Here $C$ is a constant factor with $\#$ indicating integer power due to different ways of factorization, which can be absorbed by using the fact that $||(q)_\infty||^2_S=-\frac{2\pi}{\hbar}C^2$ which gives trivial contribution in identity fusion i. e.  $||(q)_\infty||^2_{id}=1$ \cite{Beem:2012mb}. We work modulo this factor in the following content.

Before we perform the identity transformation $b\to -b$ to fuse them into partition function on $S^2\times_q S^1$, we should note that following \cite{Pasquetti:2011fj}, the holomorphic block can be alternatively regarded as the equivariant vortex partition function on $R^2 \times_q S^1$, consisting of classical, one-loop perturbative and infinite tower of non-perturbative vortex contributions. 
Interestingly, these topologically non-trivial vortex solutions arise from spontaneously breaking of gauge symmetry and exist not on Coulomb, but rather on Higgs branch.
More precisely, as also noticed in the supersymmetric partition functions defined on $S^2$ \cite{Benini:2012ui, Doroud:2012xw},  the corresponding saddle points/poles are deformations of the root of baryonic Higgs branch in the moduli space, where Coulomb and Higgs branches intersect (See \cite{Hanany:2004ea} for earlier discussions.). 
In other words, around the specific saddle points given in \eqref{FactorPole}, the partition function $Z_{S^3_b}^{\rm A}$ from one loop Coulomb branch calculation actually yields prediction for the equivariant vortex partition function on Higgs branch. In the next section, we will explicitly verify such prediction. Here for $|q|, |x|<1$, we use the Pochhammer symbol identity $\sum_{n=0}^{\infty}\frac{x^n}{(q; q)_n}=\frac{1}{(x; q)_{\infty}}$ again, rewriting \eqref{HBlock1} into:
\begin{eqnarray}\label{HBlock2}
\fB^{\rm A}{\left[\{x_i\},\{\ \by_a\}, Q_u;q\right]}\nn
&&=
\left[
\prod\limits^N_{i=1}
\theta ((q^{-1}x_i)^{\frac{1}{\sqrt{2}}};q)
\right]
\left[\prod\limits_{a=1}^{s}
\frac{\theta (\by_a x_a;q)}{\theta (\by_a;q)\theta (x_a;q)}\right]
\left[
\prod\limits^N_{p=s+1}
\left(x_p\prod\limits^s_{a=1}x_a^{-(Q_dQ_u^{-1})^a_p};q\right)_\infty
\right]_{\rm 1-loop}^{-1}\nn
\\
&&
\times
\left[
\sum\limits^\infty_{\{\hf_a=0\}}
\prod_{a=1}^s
\frac{ {y_a}^{-\hf^a}}
{ \prod \limits_{ \delta =1 }^{ (Q_u)^c_a \hf_c } (1-q^{ -\delta  }) 
}
\prod_{p=s+1}^N
\frac{1}
{\left( x_p\prod\limits^s_{a=1}x_a^{-(Q_dQ_u^{-1})^a_p};q^{-1}\right)_{(Q_d)^c_p\hf_c}}
\right]_{\rm vortex},\nn\\
\end{eqnarray}
which is more suited for matching with the vortex partition function obtained from equivariant localization.
Notice that in above for given $\{\hf_a\}$ which are interpreted as topological vortex number obtained from the quantized flux, there are all together $\sum_{a=1}^s \sum_{i=1}^N Q_{i}^a \hf_a$ terms in the denominator of vortex contribution of holomorphic block. As pointed out in \cite{Eto:2009bz}, this is precisely the dimension of the corresponding vortex moduli space with higher charge chiral fields.
Notice also that we have picked specific vacua which admit only integer vortex number while for other vacua we generally have fractional vortex number also non-trivial phase factors. It would be interesting to see how these other topological sectors can be reproduced by explicit equivariant vortex partition function computation. 

Now to proceed with the identity fusion or ``id-fusion'' mapping $b\to -b$, we re-identify the following variables: 
\begin{equation}
q=e^{\hbar}  \leftrightarrow\tilde{q}=e^{-\hbar}=q^{-1},
\quad
x_{a,p}
=q^{\frac{l_{a,p}}{2}}\eta_{a,p}  \leftrightarrow \tilde{x}_{a,p}=q^{\frac{l_{a,p}}{2}}\eta^{-1}_{a,p},
\quad
y_{a}=q^{\frac{j_{a}}{2}}\zeta_{a}  \leftrightarrow \tilde{y}_{a}=q^{\frac{j_{a}}{2}}\zeta^{-1}_{a}.
\end{equation}
Here $\{j_a\}$ and $\{l_i\}=\{l_{a}, l_p\}$ are respectively interpreted as the quantized fluxes on $S^2 \subset S^2\times_q S^1$ for the $U(1)_J^s$ topological symmetries and background $U(1)^{N}$ flavor symmetries while $\{\zeta_a\}$ and $\{\eta_i\} =\{\eta_a, \eta_p\}$ respectively are the corresponding Wilson lines along the $S^1$. 
Finally, we can apply the following identity for the theta function: 
\be
\label{id-fusion}
\left|\left| 
\theta\left(
(-q^\frac{1}{2})^bx^a;q
\right)
\right|\right|^2_{id}
=(-q^\frac{1}{2})^{-(a\cdot l)b}\eta^{-(a\cdot l)a}
\ee  
to obtain the partition function on $S^2\times_q S^1$: 
\begin{eqnarray}\label{3dIndex1}
&&{ I^{\rm A}_{3d}}(\{ \eta_i\},\{\zeta_a\} ;q)
=\sum_{\{Q_u\}}\fB^{\rm A}{\left[\{\eta_i\},\{\zeta_a\}, Q_u;q\right]}
\fB^{\rm A}{\left[\{{\eta^{-1}_i}\},\{{\zeta^{-1}_a}\}, Q_u;q^{-1}\right]}\nn
\\
&&=
\sum_{\{Q_u\}}\sum_{\{j_a\in {\mathbb Z}\}} 
\sum_{\{f_a, g_a =0\}}
\prod_{i=1}^N\prod_{a=1}^s
(-q^{\frac{l_i}{2}})\eta_i^{\frac{-l_i}{2}}
\left(\prod\limits_{b=1}^s\zeta_b^{-(Q_u^{-1})^b_a}\right)
\eta_a^{-\sum\limits^s_{b=1} j_b(Q_u^{-1})_a^b}
\nn
\\
&&
\times
\prod_{a,c=1}^s \prod_{p=s+1}^N
\frac{ 
(q^{\frac{j_c}{2}}\zeta_c       )^{ -(Q_u^{-1})^c_a f^a} 
(q^{\frac{j_c}{2}}\zeta_c^{-1})^{-(Q_u^{-1})^c_a g^a}
}
{
{ \prod_{ \delta =1 }^{ f_a }{ (1-q^{ -\delta  }) }  }
\prod^{g_a}_{\Delta=1}(1-q^{\Delta}) 
\left[
\left(
q^{l_p/2} \eta_p(q^{l_a/2+f_a}\eta_a)^{-(Q_dQ_u^{-1})^a_p};q
\right)_{\infty}
\left(
q^{l_p/2} \eta_p^{-1}(q^{l_a/2-g_a}\eta^{-1}_a)^{-(Q_dQ_u^{-1})^a_p};q^{-1}
\right)_{\infty} 
\right]
} .
\nonumber \\
\end{eqnarray}
Here the summation over $\{f_a, g_a\}$ indicates that we are fusing all the different holomorphic blocks associated with different sectors in \eqref{FactorPole} via identity fusion to obtain  \eqref{3dIndex1}, for which the resultant expression should therefore be regarded as the integrated expression for the partition function on $S^2\times_q S^1$ along the contour which picks up all the simple poles in the integrand.

To interpolate to the partition function on $S^2$, we can restore the $S^1$ radius dependence in the deformation parameter $q$ and take the limit $q\to 1^{-}$ while keeping fixed the appropriate masses and FI parameters for making the identifications with those in the two dimensional $\cN=(2,2)$ gauged linear sigma models. 
The resultant gauged linear sigma model provides the low energy effective description on the Higgs branches of Theory A upon compactification on a circle \cite{Aganagic:2001uw} and echoes the interpretation of holomorphic block as the K-theoretic vortex partition function.  We will briefly sketch out this interpolation in the next section. 
In addition, we will also take an alternative and perhaps more illuminating route to two dimensions, by first rewriting the integrands of the ellipsoid partition function into manifestly factorizable form, and perform the fusion procedure before the contour integration. 
Along the way, we demonstrate how we can directly recover the $S^2$ partition function for the $\cN=(2,2)$ Landau-Ginzburg model, which on the other hand gives the low energy effective description on the Coulomb branch of compactified Theory A.
These two different routes explicitly illustrate that 3d mirror symmetry exchanging Higgs and Coulomb branches, indeed interpolates to 2d mirror symmetry exchanging GLSM and LG model.

\subsubsection{GLSM and Landau-Ginzburg Twisted Superpotential}
\paragraph{}
Using the results in \cite{Beem:2012mb} which states that with appropriate choices of integration contour, the factorization property of the partition function can commute with the integration, we can rewrite $Z_{S^3_b}^{\rm A}(m_i, \zeta^a)$ \eqref{integrand ThA} into:
\begin{eqnarray}\label{integrand2 ThA}
Z^{\rm A}_{S^3_b}(m_i; \zeta^a)
&=&
\int d^s\sigma 
\prod\limits^N_{i=1}
\frac{e^{i\Phi}||{B_\Delta}_i||^2_S}
{\left(\left|\left|
\theta \left(((-q^{\frac{1}{2}})^{-\sqrt{2}})x_i^{\frac{1}{\sqrt{2}}};q\right)\right|\right|^2_S\right)
\left(
i^\#C^\#\left|\left|\theta 
\left((-q^{-\frac{1}{2}})\prod\limits^s_{a=1}s_a^{Q^a_i};q\right)
\right|\right|^2_S
\right)}
\prod\limits^s_{a=1}
\left|\left|
\frac{\theta \left(\prod\limits^N_{i=1}  x_i^\frac{Q^a_i}{2}e^{2\pi b\zeta^a} ;q\right)\theta \left(s_a ;q\right)}
{\theta \left(\prod\limits^N_{i=1} x_i^\frac{Q^a_i}{2}e^{2\pi b\zeta^a}s_a ;q\right)}
\right|\right|^2_S,\nn\\
\end{eqnarray}
where we have defined the following variables:
\bea
&&q=e^{2\pi ib^2},\quad s_a=e^{2\pi b \sigma^a},\quad x_i=e^{2\pi b m_i}, \ \quad a=1,\dots, s, \quad i=1,\dots, N\nn\\
&&{B_\Delta}_i = \left( \prod_{a=1}^s s_a^{-Q^a_i}x^{-1}_iq;q\right)_\infty, \quad
e^{i\Upphi}=\exp\left(
\frac{-i\pi}{2}Nc_b^2
\right).
\eea
In expressing the integrand in \eqref{integrand2 ThA} into manifestly factorizable form, we have also packaged the FI and Chern-Simons contributions in terms of theta functions defined in \eqref{theta function}\footnote{In following discussion, we can also consider the background Chern-Simons coupling. However, the cancellation of anomalous term is not manifest in the integration form, so we just keep it in mind but not put it in the calculation.}.
Given \eqref{integrand2 ThA}, we can now directly change from $S$-fusion to identity fusion within the integrand, 
after re-defining the following parameters: $s_a =q^\frac{n_a}{2}t_a$ , $\tilde{s}_a =q^\frac{n_a}{2}t_a^{-1} $, $x_i=q^\frac{l_i}{2}\eta_i$ , $\tilde{x}_i =q^\frac{l_i}{2}\eta_i^{-1} $, $e^{2\pi b\zeta^a}=q^\frac{j_a}{2}p_a $ ,  $\tilde{ e^{2\pi b\zeta^a}}=q^\frac{j_a}{2}p_a^{-1} $.
The partition function on $S^2\times_q S^1$ can now be written as: 
\begin{eqnarray}
\label{index basic form}
I^{\rm A}_{3d}
=&&
\sum\limits_{n_a\in Z^s}
\int \prod\limits^s_{a=1}\left( \frac{dt_a}{2\pi i t_a} \right) 
\left(
\prod\limits^N_{i=1}
(-q^{\frac{1}{2}})^{l_i}\eta_i^{\frac{-l_i}{2}}\right)^{-1}
\prod\limits^N_{i=1}
\left(
\prod\limits^\infty_{r=0}
\frac
{1-q^{r+1}
\left(\prod\limits^s_{a=1} (q^{\frac{n_a}{2}}t_a)^{-Q^a_i}\right)q^{\frac{-l_i}{2}}\eta_i^{-1}}
{1-q^r\left(\prod\limits^s_{a=1} (q^{\frac{n_a}{2}}t_a^{-1})^{-Q^a_i}\right)q^{\frac{-l_i}{2}}\eta_i}
\right)
\nn\\
\times&&
\prod^N_{i=1}\left(
(-q^\frac{1}{2})^{(\sum_a Q^a_i\cdot n_a)}
\prod\limits^s_{c=1} t_c^{-(\sum_a Q^a_i\cdot n_a)Q^c_i}
\right)^{-1}
\prod\limits^s_{a=1}
\left(
\prod\limits^N_{i=1}
\eta_i^{n_a\frac{Q^a_i}{2}}
p_a^{n_a}
t_a^{(\sum_i\frac{Q^a_i}{2}\cdot l_i +j_a)}
\right).
\end{eqnarray}
Let us pause here to comment on how $I_{3d}^{\rm A}$ can be related to the corresponding 2d GLSM theory. 
Firstly, the infinite product coming from the chiral multiplets can be written into ratio of q-deformed gamma functions which reduce to the ratio of ordinary gamma functions when taking the $q\to 1^{-}$. 
This is precisely the form of chiral multiplets contribution to the $S^2$ partition function in 2d up to a divergent pre-factor \cite{Fitouhi, Rahman}\footnote{This prefactor gives a non-trivial rescaling associated to a factor of topological $U(1)$ of gauge symmetry, discussed in \cite{Benini:2012ui}.}:
\be
\prod\limits^\infty_{r=0}
\frac
{1-q^{r+1}\left(\prod\limits^s_{a=1} (q^{\frac{n_a}{2}}t_a)^{-Q^a_i}\right)q^{\frac{-l_i}{2}}\eta^{-1}_i}
{1-q^r\left(\prod\limits^s_{a=1} (q^{\frac{n_a}{2}}t_a^{-1})^{-Q^a_i}\right)q^{\frac{-l_i}{2}}\eta_i}
=
(1-q)^{-1+2(\log_q\eta_i+\sum^s_{a=1}Q^a_i \log_q t_a)}
\frac{\Gamma_q\left(
-\sum^s_{a=1}\frac{n_aQ_i^a}{2}-\frac{l_i}{2} +\log_q\eta_i+\sum^s_{a=1}Q^a_i \log_q t_a\right)}{\Gamma_q\left(
1-\sum^s_{a=1}\frac{n_aQ_i^a}{2}-\frac{l_i}{2} -\log_q\eta_i-\sum^s_{a=1}Q^a_i \log_q t_a\right)},
\ee
where we have also introduced the q-deformed Gamma function $\Gamma_q(x)$:
\be
\Gamma_q(x)=(1-q)^{1-x}\prod^{\infty}_{r=0} \frac{1-q^{r+1}}{1-q^{r+x}}=(1-q)^{1-x}\frac{(q;q)_{\infty}}{(q^x;q)_{\infty}}.
\ee 
The parameter identifications are given as following. By comparing with the contribution of chiral multiplets in the equation (3.35) of \cite{Benini:2012ui}. $Q^a_i$ remains the charge of i-th chiral multiplet under $U(1)_a$ gauge group; $\lim\limits_{q\rightarrow 1^-}\log_q t_a$ plays the lowest component of 2d vector multiplet;  $n_a$ is the quantized magnetic flux on $S^2$
while the remaining $\lim\limits_{q\rightarrow 1^-}\log_q \eta_i$ and $l_i$ combine to form the complex mass parameters. 
Moreover, the first term and third term in \eqref{index basic form} combine to give mass or gauge field square term derived from 3d Chern-Simons term while in the last term $\sum_{i=1}^N Q^a_il_i/2+j_a$ plays renormalized FI parameter, and $\lim\limits_{q\rightarrow 1^-}\left(\log_qp_a+\sum_{i=1}^NQ^a_i\log_q\eta_i/2\right)$ plays the 2d topological $\theta$ angle.

We now demonstrate that, in the small radius limit, the 3d-index $I_{3d}^{\rm A}$ can also be dimensionally reduced into Landau-Ginzburg model which is mirror dual to the above GLSM. Let us first relate the asymptotic behavior of infinite Pochhammer symbol to the dilogarithm function \cite{Eynard} $(qz; q)_{\infty} \xlongrightarrow{q\rightarrow 1^-}\exp \left(\frac{1}{\log q}\sum_{m=0}^{\infty}\frac{B_m (\log q)^m}{m !}{\rm Li}_{2-m}(z)\right)$ \footnote{There the $B_m=\left( 1, \frac{1}{2},\frac{1}{6}, 0, -\frac{1}{30}, \cdots\right)$ is the $n^{th}$ Bernoulli number.}, rewrite \eqref{index basic form} into the following form, pick out the lowest term, apply a reflection relation of dilogarithm, ${\rm Li_2}(x)+{\rm Li_2(x^{-1})}=-1/6 \pi^2 -1/2 [\ln (-z)]^2$, and get \footnote{That $\prod^N_{i=1}\eta_i^{\frac{-l_i}{2}}$ can be cancelled by considering the background flavor Chern-Simons term.} :
\begin{eqnarray}
I_{3d}^{\rm A}
&\sim& 
\int \prod\limits^s_{a=1}\left( \frac{dt_a}{2\pi i t_a} \right) 
(-1)^{\frac{-1}{2}\sum_in_aQ^a_i+l_i+2}
\left(\prod\limits^N_{i=1}
\eta_i^{\frac{-l_i}{2}}\right)
\prod\limits^s_{a=1}
\left(
\prod\limits^N_{i=1}
\eta_i^{-n_a\frac{Q^a_i}{2}}
p_a^{n_a}
t_a^{(\sum_iQ^a_i-\frac{Q^a_i}{2}\cdot l_i +j_a)}
\right)
\nonumber\\
&\times&
\prod\limits^N_{i=1}
\left(
\exp \left(
\frac{1}{\log q}
\left[
{\rm Li}_{2}
\left(
(\eta_i\prod_at_a^{Q^a_i})^{-1}
q^{\frac{-1}{2}(\sum_an_aQ^a_i+l_i+2)}
\right)
-
{\rm Li}_{2}
\left(
(\eta_i\prod_at_a^{Q^a_i})
q^{\frac{-1}{2}(\sum_an_aQ^a_i+l_i)}
\right)
\right]
\right)
\right)
\nonumber\\
\end{eqnarray}
As we take 2d limit $q \to 1^{-} $, the summation expansion in the exponential factors of Pochhammer symbol will leave only the lowest term, i.e. m=0, the dilogarithm $\rm Li_2(x)$ contributions. Such contributions can be viewed as coming from integrating the KK modes giving the superpotential terms as \cite{Aganagic:2001uw} 
\begin{equation}\label{DeltaW}
\Delta \tilde{W}
=-\sum^N_{i=1}\sum_{n\in \mathbb{Z}}
\left(\Sigma_i+\frac{in}{R}\right)
\left[\log\left( \Sigma_i+\frac{in}{R}\right) -1\right]
=\frac{1}{2\pi R}
\left[ 
\sum\limits^N_{i=1}
-{\rm Li_2}(e^{-2\pi R \Sigma_i})
+\frac{1}{4}\left(2\pi R \Sigma_i\right)^2
\right]
\end{equation}
where
\begin{equation}\label{Sigmaa}
\Sigma_i=\sum\limits^s_{a=1}Q_i^a\Sigma_a+m_i,
\end{equation}
and $\Sigma_a$ is the field strength of the vector field. The square terms are cancelled by the Chern-Simons and FI contributions, and the remaining term precisely gives the Landau-Ginzburg superpotential. 

At the finite radius, we can also represent the index \eqref{index basic form} in terms of the integral for q-deformed Bessel functions.
Let us first recall the following identity between q-deformed version of Gamma and Bessel functions  \cite{Fitouhi}\cite{Rahman}\cite{Koornwinder}: 
\be\label{qdeform-ID}
\frac{(1+q)^t\Gamma_{q^2}((\a +1+t)/2)}{\Gamma_{q^2}((\a +1-t)/2)}=\int\limits^\infty_o x^tJ_\a ((1-q)x;q^2)d_qx ,
\ee
where the q-integration is defined as
\be
\int\limits^\infty_0f(x)d_qx=(1-q)\sum\limits^\infty_{k=-\infty}f(q^k)q^k ,
\ee
and the q-Bessel function is given by  
\be
J_\a (x;q^2)=\Gamma_{q^2}(\a +1)\sum_{n=0}^\infty(-1)^n\frac{q^{n(n-1)}}{\Gamma_{q^2}(\a +n+1)\Gamma_{q^2}(n+1)}\left(\frac{x}{1+q}\right)^{2n}.
\ee
In the $q\rightarrow 1^-$ limit, the q-defomed Gamma and Bessel functions,  $\Gamma_{q^2}(x)$ and $J_{\alpha}(x; q^2)$ respectively, reduce to the ordinary Gamma and Bessel functions, $\Gamma(x)$ and $J_{\alpha}(x)$, while the identity \eqref{qdeform-ID} reduces to the Fourier transform of Bessel functions $J_{\alpha}(x)$.
Finally collecting the pieces, we can recast the 3d index \eqref{index basic form} into:
\begin{eqnarray}\label{NewIndex3D}
I^{\rm A}_{3d}&\sim &
\sum\limits_{n_a \in {\mathbb Z}^s}
\int\prod\limits^s_{a=1}\left( \frac{dt_a}{2\pi i t_a} \right)
\left(\prod\limits^N_{i=1}
(-q^{\frac{1}{2}})^{l_i}\eta_i^{\frac{-l_i}{2}}\right)^{-1}
\prod\limits^N_{i=1}
\left(
\Delta^{\b_i}\int\limits^\infty_0
e^{\b_i\tau_i}
J_{\a_i}(\Delta^{\tau_i};q)d_{\qq}e^{\tau_i}
\right)
\nn\\
&\times&
\prod^N_{i=1}\left(
(-q^\frac{1}{2})^{(\sum_a Q^a_i n_a)}
\prod\limits^s_{c=1}
t_c^{-(\sum_a Q^a_i n_a)Q^c_i}
\right)^{-1}
\prod\limits^s_{a=1}
\left(
\prod\limits^N_{i=1}
\eta_i^{n_a\frac{Q^a_i}{2}}
p_a^{n_a}
t_a^{(\sum_i\frac{Q^a_i}{2} m_i +l_a)}
\right)
\nn\\
&=&\sum\limits_{n_a \in {\mathbb Z}^s}
\int \prod\limits^s_{a=1}\left( \frac{dt_a}{2\pi i t_a} \right)
\left(\prod\limits^N_{i=1}
(-q^{\frac{1}{2}})^{l_i}\eta_i^{\frac{-l_i}{2}}\right)^{-1}
\int\limits^\infty_0
\prod\limits^N_{i=1}
\left(
d_{\qq}e^{\tau_i}\right)
\int\limits^\pi_{-\pi}
\left(\prod\limits^N_{i=1}
\frac{d\phi_i}{2\pi }
\right)
\prod\limits^N_{i=1}
\left(
\Delta^{\b_i}
e^{\b_i\tau_i}e^{-i\a_i\phi_i}\right)
\nn\\
&\times&\prod\limits^N_{i=1}
\exp \left[
\frac{1}{\log q}
\sum\limits^\infty_{m=0}\frac{B_m}{m!}(\log q)^m
\left(
{\rm Li}_{2-m}(\frac{-\Delta^2 e^{2\tau_i}}{q})
-{\rm Li}_{2-m}(\frac{\Delta e^{\tau_i}e^{i\a_i\phi_i}}{q})
-{\rm Li}_{2-m}(\frac{-\Delta e^{2\tau_i}e^{-i\a_i\phi_i}}{q})
\right)
\right]
\nn\\
&\times&
\prod^N_{i=1}\left(
(-q^\frac{1}{2})^{(\sum_a Q^a_i n_a)}
\prod\limits^s_{c=1}
t_c^{-(\sum_a Q^a_i n_a)Q^c_i}
\right)^{-1}
\prod\limits^s_{a=1}
\left(
\prod\limits^N_{i=1}
\eta_i^{n_a\frac{Q^a_i}{2}}
p_a^{n_a}
t_a^{(\sum_i\frac{Q^a_i}{2}m_i +l_a)}
\right)
\end{eqnarray}
with the variables, 
\be
\left(\prod\limits^s_{a=1} (q^{\frac{n_a}{2}}t_a)^{-Q^a_i}\right)q^{\frac{-l_i}{2}}\eta^{-1}_i 
\equiv q^{\frac{\a_i -1-\b_i}{2}}, \quad
\left(\prod\limits^s_{a=1} (q^{\frac{n_a}{2}}t_a^{-1})^{-Q^a_i}\right)q^{\frac{-l_i}{2}}\eta_i
\equiv q^{\frac{\a_i+1+\b_i}{2}},\quad
\Delta \equiv 1-q^{\frac{1}{2}}.
\ee
To obtain the second equality in \eqref{NewIndex3D}, we have further expanded the q-deformed Bessel functions, by
\be
J_\a(x;q)=\frac{1}{2\pi}\int\limits^\pi_{-\pi}
d\theta 
e^{-i\a \theta}
\frac{(-x^2;q)_\infty}{(xe^{i\a\theta};q)_\infty(-xe^{-i\a\theta};q)_\infty}.
\ee
In doing so, when taking the limits $q\rightarrow 1^{-}$, we make the q-deformed Bessel function $J_{\a_i}(x;q)\rightarrow J_{\a_i}(x)$.
As a result, the theory reduces to the $S^2$ partition function for $\cN=(2,2)$ Landau-Ginzburg model obtained in \cite{Gomis:2012wy}.
Finally, we have shown this 3d index as 2d theory accompanied by KK-modes contributions.
To sum up, we show our original 3d mirror symmetry, if applying above discussion on Theory A and Theory B respectively, can be considered as the equivalence between 2d GLSM plus certain q-deformation contributions of one theory and the Landau-Ginzburg theory with KK-modes running around the additional  $S^1$ of the other theory.

\subsection{Comments on Non-Abelian Gauge Generalization}
\paragraph{}
In this section, we will make the initial steps to generalize our previous analysis to the non-abelian theories, where the main additions come from the non-Abelian vector multiplet contributions. Here we consider three dimensional $\cN=2$  $U(N)$ gauge theory with $N_f$ chiral multiplets and $N_f$ anti-chiral multiplets. 
We also turn on the FI parameter $\xi$ and the complex mass parameters $\{m_i, \hat{m}_i\}$ for the chiral and anti-chiral multiplets. We would like to note that the three dimensional mirror dual of this example was first derived in \cite{deBoer:1997ka} using D-brane construction and $SL(2,Z)$ duality in IIB string theory, where the resultant theory contains quiver gauge group and chiral multiplets in various representations. In addition to the tree-level superpotential, there are dynamically generated non-perturbative corrections due to three dimensional instantons or monopoles, rendering the explicit check of mirror symmetry difficult.  
However, assuming three dimensional non-Abelian mirror symmetry persists under these non-perturbative corrections, the interpolation procedure using localization earlier can also be applied to derive various new non-Abelian mirror pairs in two dimensions.
The factorization of partition function for this example has been carried out in \cite{Taki:2013opa}, where the author showed that we can decompose it into holomorphic and anti-holomorphic parts. With the factorizability, we then glue them into partition function on $S^2\times_q S^1$ for interpolation.
We start with the partition function: \footnote{The same argument for background Chern-Simons coupling can applies here to canceling the Chern-Simons-like contribution in \eqref{Non-ablian partition function}.}  
\begin{eqnarray}
\label{Non-ablian partition function}
Z_{S^3_b}^{\rm U(N)}&=&
\int \frac{d^N \sigma}{N!} e^{-2\pi i\xi\sum_\a\sigma_\a}
\pl_{1\leq \a <\b \leq N}
4\sinh (-\pi b (\sigma_\a-\sigma_\b))\sinh (-\pi b^{-1} (\sigma_\a-\sigma_\b))
\pl^N_{\a=1}\pl^{N_f}_{i=1}
\frac{s_b(c_b-\sigma_\a-m_i)}
     {s_b(-c_b-\sigma_\a-\m_i)}
\nn\\
&=&
\int \frac{d^N \sigma}{N!}
 e^{\frac{i\pi}{2}\sum\limits^{N_f}_{i=1}(-m_i^2+\m_i^2+2c_b(m_i+\m_i))N}
\pl_{1\leq \a <\b \leq N}
\( e^{-\pi b (\sigma_\a-\sigma_\b)}-e^{\pi b (\sigma_\a-\sigma_\b)}\)
\( e^{-\pi b^{-1} (\sigma_\a-\sigma_\b)}-e^{\pi b^{-1} (\sigma_\a-\sigma_\b)}\)
\nn\\
&\times&
\frac{1}
{\exp\left[2\pi i\(\xi-c_bN_f+\frac{\sum_im_i-\m_i}{2}\)\sum\limits_{\a =1}^N\sigma_\a \right]}
\pl^N_{\a=1}
\pl^{N_f}_{i=1}
\frac{\( e^{2\pi (2c_b-\sigma_\a-m_i)b     };e^{2i\pi b^2}    \)_\infty}
     {\( e^{2\pi (    -\sigma_\a-m_i)b^{-1}};e^{-2i\pi b^{-2}}\)_\infty}
\frac{\( e^{2\pi (-\sigma_\a-\m_i-2c_b)b^{-1}};e^{-2i\pi b^{-2}}\)_\infty}
     {\( e^{2\pi (    -\sigma_\a-\m_i)b     };e^{ 2i\pi b^{ 2}}\)_\infty}.\nn\\
\end{eqnarray}
By identifying the holomorphic parameters: 
\begin{equation}
s_\a=e^{2\pi b \sigma_\a},A_i=e^{2\pi b m_i},B_i=e^{2\pi b \m_i},C =e^{2\pi b \xi}
\end{equation}
as well as the anti-holomorphic counterpart changing $b\rightarrow b^{-1}$,
we can rewrite them as $S$-fusion:
\begin{eqnarray}
&&Z^{\rm U(N)}_{S^3_b}=
\frac{1}{N!}\int d^N\sigma 
\prod\limits^{N_f}_{i=1}
\left(\left|\left|
\theta ((-q^\frac{1}{2})^{-\sqrt{2}}A_i^{\frac{1}{\sqrt{2}}};q)\right|\right|^2_{S}\right)^N
\prod\limits^{N_f}_{i=1}
\left(\left|\left|
\theta ((-q^\frac{1}{2})^{-\sqrt{2}}B_i^{\frac{1}{\sqrt{2}}};q)\right|\right|^2_{S}\right)^{-N}
\nn\\
&&
\times
\pl_{1\leq \a <\b \leq N}
\( s_\a^\0.5  s_\b^{-\0.5}- s_\a^{-\0.5}  s_\b^{\0.5}\)
\(\s_\a^\0.5 \s_\b^{-\0.5}-\s_\a^{-\0.5} \s_\b^{\0.5}\)
\pl_{\a =1}^N\pl_{i=1}^{N_f}\( \l||  \frac{\theta\( C\pl_{i=1}^{N_f}(A_iB_i^{-1})^\0.5;q\)\theta\(s_\a ;q\)}{\theta\( C\pl_{i=1}^{N_f}(A_iB_i^{-1})^\0.5s_\a;q\)}  \r||^2_S
\frac{(s_\a^{-1} A_i^{-1}q;q)_\infty}{(\s_\a^{-1}\tilde{A}_i^{-1};\q)_\infty}
\frac{(\s_\a^{-1}\tilde{B}_i^{-1}\q;\q)_\infty}{(s_\a ^{-1}B_i^{-1} ;q)_\infty}\).
\nn\\
\end{eqnarray}
Notice that contrasting with \cite{Taki:2013opa}, we have also factorized the classical piece as well as the perturbative and non-perturbative pieces.
To get the 3d index, we should identify the index parameters:
\be
s_\a =q^{\frac{n_\a}{2}}t_\a,\quad\s_\a =q^{\frac{n_\a}{2}}t_\a^{-1},
\quad A_i=q^{\frac{m_i}{2}}\zeta_i,\quad B_i=q^{\frac{\m_i}{2}}\eta_i,\quad C =q^{\frac{l}{2}}p, 
\ee
and, by $id$-fusion, we get the 3d index :
\begin{eqnarray}
I^{\rm U(N)}&=&\sum_{n_\a \in { Z}}\frac{1}{N!} \int
\frac{d^N t^\a}{2\pi i t^\a} 
\left(\prod\limits^N_{i=1}
(-q^{\frac{1}{2}})^{m_i}\zeta_i^{\frac{-m_i}{2}}\right)^N
\left(\prod\limits^N_{i=1}
(-q^{\frac{1}{2}})^{\hat{m}_i}\eta_i^{\frac{-\hat{m}_i}{2}}\right)^{-N}
\pl_{1\leq \a <\b \leq N}
\(q^{\frac{n_\a -n_\b}{2}}+q^{-\frac{n_\a -n_\b}{2}}-t_\a t_\b^{-1}-t_\a^{-1}t_\b\)
\nn\\
&\times&
\pl_{\a =1}^N \( t_\a^{\(l +\sum\limits^{N_f}_{i=1}(\frac{m_i}{2}-\frac{\m_i}{2})\)}
\(\pl_{i=1}^{N_f} p\zeta^\0.5_i\eta^{-\0.5}_i\)^{n_\a}\)
\nn\\
&\times&
\pl^N_{\a =1}\pl^{N_f}_{i=1}
\frac{\Gamma_q\(-\frac{n_\a}{2}-\frac{m_i}{2}+\log_q(t_\a \zeta_i)\)}
     {\Gamma_q\(1-\frac{n_\a}{2}-\frac{m_i}{2}-\log_q(t_\a \zeta_i)\)}
\frac{\Gamma_q\(-\frac{n_\a}{2}-\frac{\m_i}{2}-\log_q(t_\a \eta_i)\)}
     {\Gamma_q\(1-\frac{n_\a}{2}-\frac{\m_i}{2}+\log_q(t_\a \eta_i)\)}
\frac{1}{(1-q)^{2-2\log_q\zeta_i+2\log_q\eta_i}}.\nn\\
\end{eqnarray}
By writing them into q-deformed Gamma functions, it becomes obvious that in the limit $q\rightarrow 1^-$, this index will reduce to the $S^2$ partition function for GLSM in discussed in \cite{Gomis:2012wy}.
Furthermore, we can change them into the q-deformed Bessel function and expand their Fourier modes:
\begin{eqnarray}
&&I^{\rm U(N)}
=\sum_{n_\a \in { Z}}\frac{1}{N!} \int
\frac{d^N t^\a}{2\pi i t^\a} 
\left(\prod\limits^N_{i=1}
(-q^{\frac{1}{2}})^{m_i}\zeta_i^{\frac{-m_i}{2}}\right)^N
\left(\prod\limits^N_{i=1}
(-q^{\frac{1}{2}})^{\hat{m}_i}\eta_i^{\frac{-\hat{m}_i}{2}}\right)^{-N}
\int\limits^\infty_0
\int\limits^\pi_{- \pi}
\prod\limits^N_\a
\prod\limits^{N_f}_i
\frac{ d\phi_{\a i}}{2\pi} d_{q^\0.5}e^{\tau^\a_i}
\int\limits^\infty_0
\int\limits^\pi_{- \pi}
\prod\limits^N_\a
\prod\limits^{N_f}_i
\frac{ d\phii_{\a i}}{2\pi} d_{q^\0.5}e^{\tilde{\tau}^\a_i}
\nn\\
&\times&
\pl_{1\leq \a <\b \leq N}
\(q^{\frac{n_\a -n_\b}{2}}+q^{-\frac{n_\a -n_\b}{2}}-t_\a t_\b^{-1}-t_\a^{-1}t_\b\)
\pl_{\a =1}^N \( t_\a^{\(l +\sum\limits^{N_f}_{i=1}(\frac{m_i}{2}-\frac{\m_i}{2})\)}
\(\pl_{i=1}^{N_f} p\zeta^\0.5_i\eta^{-\0.5}_i\)^{n_\a}\)
\(\pl^N_{\a =1}\pl^{N_f}_{i=1}
\Delta^{\omega_{\a i}+\tilde{\omega}_{\a i}}\)
\nn\\
&\times& 
\pl^N_{\a =1}\pl^{N_f}_{i=1}
\left[
e^{\tau^\a_i\omega_{\a i}-i\chi_{\a i}\phi_{\a i}}
\right.
\exp\left\{\sum\limits^\infty_{m=0}\frac{B_m}{m!}(\log q)^{m-1}
\left[{\rm Li}_{2-m}\(\frac{-\Delta^2e^{2\tau^\a_i}}{q}\)
\left.
-{\rm Li}_{2-m}\(\frac{\Delta e^{\tau^\a_i}e^{i\chi_{\a i}\phi_{\a i}}}{q}\)
-{\rm Li}_{2-m}\(\frac{-\Delta e^{\tau^\a_i}e^{-i\chi_{\a i}\phi_{\a i}}}{q}\)
\right]
\right\}\right]
\nn\\
&\times&
\pl^N_{\a =1}\pl^{N_f}_{i=1}
\left[
e^{\tilde{\tau}^\a_i\tilde{\omega}^\a_i-i\tilde{\chi}_{\a i}\phii_{\a i}}
\right.
\exp\left\{\sum\limits^\infty_{m=0}\frac{B_m}{m!}(\log q)^{m-1}
\left[{\rm Li}_{2-m}\(\frac{-\Delta^2e^{2\tilde{\tau}^\a_i}}{q}\)
\left.
-{\rm Li}_{2-m}\(\frac{\Delta e^{\tilde{\tau}^\a_i}e^{i\tilde{\chi}_{\a i}\phii_{\a i}}}{q}\)
-{\rm Li}_{2-m}\(\frac{-\Delta e^{\tilde{\tau}^\a_i}e^{-i\tilde{\chi}_{\a i}\phii_{\a i}}}{q}\)
\right]
\right\}\right].\nn
\\
\end{eqnarray}
The final expression is then the Landau-Ginzburg superpotential expanded as the polynomial of the radius of $S^1$. For the same purpose, taking limit to 2d, it will reduce to the non-abelian case in \cite{Gomis:2012wy}.

\section{Equivariant Localization and Vortex Partition Function}\label{EquivLoc}
\paragraph{}
In \cite{Doroud:2012xw}, it derives the supersymmetric fixed point equations on $S^2$ and showed that the vortices are localized on the north and south poles by introducing a $\mathcal{Q}$-exact deformation term. We expect there will be similar fixed points corresponding to vortices on ellipsoid $S^3_b$. The Heegaard splitting tells us that the ellipsoid $S^3_b$ can be factorized into two components, i.e. two solid tori $D^2 \times_q S^1$ and $D^2\times_{\tilde{q}} S^1$ with the fugacities, $q=\mathrm{exp}(2\pi i b^2)$ and $\q=\mathrm{exp}(2 \pi i b^{-2})$ respectively. The boundaries of the solid tori are identified by the $S$ element of the $SL(2,\mathbb{Z})$, together with a reversal of orientation.

\subsection{Supersymmetric Fixed Point Equations on Ellipsoid $S_b^3$}
\paragraph{}
The ellipsoid $S_b^3$ preserves $U(1) \times U(1)$ global symmetries and its metric can be written in terms of the embedding coordinates as \cite{Hama:2011ea}:
\begin{equation}
\label{squashed 3 sphere metric}
ds^2 = l^2(dx_0^2+dx_1^2)+ \tilde{l}^2(dx_2^2+dx_3^2), \quad x_0^2+x_1^2+x_2^2+x_3^2=1.
\end{equation}
Inserting $(x_0,x_1,x_2,x_3)=(\mathrm{cos}\theta\mathrm{cos}\varphi,\mathrm{cos}\theta\mathrm{sin}\varphi,\mathrm{sin}\theta\mathrm{cos}\chi,\mathrm{sin}\theta\mathrm{sin}\chi)$, one obtains
\begin{equation}
\label{squashed 3 sphere metric in sph-coor}
ds^2 = f(\theta)^2 d\theta^2+l^2\mathrm{cos}^2\theta d\varphi^2+\tilde{l}^2\mathrm{sin}^2\theta d\chi^2, \quad
 f(\theta)=\sqrt{l^2\mathrm{sin}^2\theta+\tilde{l}^2\mathrm{cos}^2\theta}.
\end{equation}
We now derive the supersymmetric fixed point equations by demanding the supersymmetric variations for fermions to vanish:
\begin{equation}
\delta \lambda=0, \quad \delta \bar{\lambda}=0, \quad \delta \psi=0, \quad \delta \bar{\psi}=0,
\end{equation}
where $\lambda,\bar{\lambda}$ are fermions in vector multiplet and $\psi,\bar{\psi}$ are fermions in chiral multiplet.
The explicit supersymmetry variations are given in Appendix \ref{SqS3SUSY}. 
Substituting the killing spinors,
\begin{equation}
\epsilon =\psi_{+-}=
\begin{pmatrix}
e^{\frac{i}{2} (\chi-\varphi+\theta)}\\
-e^{\frac{i}{2} (\chi-\varphi-\theta)}
\end{pmatrix}, \quad
\bar{\epsilon} =\psi_{-+}=
\begin{pmatrix}
e^{\frac{i}{2} (\chi-\varphi+\theta)}\\
e^{\frac{i}{2} (\chi-\varphi-\theta)}
\end{pmatrix},
\end{equation}
into the SUSY transformation \eqref{SUSY trans for vector fermion} and \eqref{SUSY trans for chiral fermion}, one gets after some manipulations: 
\begin{align}
& \mathrm{cos}\frac{\theta}{2}(-D_2 \sigma + 2iF_{12})+i\mathrm{sin}\frac{\theta}{2}(iD_1 \sigma)=0, \\
& i\mathrm{sin}\frac{\theta}{2}(D_2 \sigma + 2iF_{12}) + \mathrm{cos}\frac{\theta}{2}(-iD_1 \sigma)=0, \\
\label{conf_1}
& \mathrm{cos}\frac{\theta}{2}(-\frac{2}{3}iQ\sigma V_2 + iD_3 \sigma +2iF_{31})+i\mathrm{sin}\frac{\theta}{2}(-\frac{2}{3}Q\sigma V_1-\frac{Q\sigma}{f}-D+2iF_{23})=0, \\
\label{conf_2}
& i\mathrm{sin}\frac{\theta}{2} (\frac{2}{3}iQ\sigma V_2 + iD_3 \sigma -2iF_{31}) + \mathrm{cos}\frac{\theta}{2}(\frac{2}{3}Q\sigma V_1-\frac{Q\sigma}{f}-D-2iF_{23}) =0, \\
& \mathrm{cos}\frac{\theta}{2} (-D_2\phi + iD_3 \phi -\frac{2}{3} \phi V_2)+ i\mathrm{sin}\frac{\theta}{2} (i D_1 \phi - \frac{2}{3}\phi V_1 -\frac{\phi}{f})=0,\\
& i\mathrm{sin}\frac{\theta}{2} (D_2\phi + iD_3 \phi +\frac{2}{3} \phi V_2) + \mathrm{cos}\frac{\theta}{2} (-i D_1 \phi + \frac{2}{3}\phi V_1 -\frac{\phi}{f})=0, \\
& F=0, \\
&Q \sigma \phi=0,
\end{align}
where
\begin{equation}
V_1 =-\frac{1}{2}(1-{l}/{f(\theta)}), \quad
V_2 =\frac{1}{2}(1-{\tilde{l}}/{f(\theta)}), \quad
b^2 = \frac{\tilde{l}}{l}
\end{equation}
and $\phi$ is the scalar in chiral multiplet of charge $Q$; $\sigma$ is the scalar in vector multiplet; $F$ and $D$ are auxiliary fields.
One can also include the twisted mass for the chiral multiplet by simply replacing
$Q\sigma$ by $Q\sigma+m$. 

We can also introduce the following parameters:
\begin{equation}\label{newcoord1}
y_{0,1}=l x_{0,1}; \quad y_{2,3}=\tilde{l} x_{2,3};\quad  l=\frac{\rho}{b}; \quad \tilde{l}=\rho b,
\end{equation} 
where $\rho$ has the dimension of length, and we can deduce the relations:
\be\label{newcoord2}
b\frac{\sqrt{y_0^2+y_1^2}}{\rho} =\cos\theta, \quad \frac{1}{b}\frac{\sqrt{y_2^2+y_3^2}}{\rho}=\sin\theta.
\ee
Now if we set $\theta \to 0$, from \eqref{newcoord2} this corresponds to the limit $b\to \infty$ while taking $\sqrt{y_0^2+y_1^2}/\rho =\sqrt{x_0^2+x_1^2}/b \to 0$ and keeping $\sqrt{y_2^2+y_3^2}/\rho = b \sqrt{x_{2}^2+x_3^2}$ fixed and arbitrary, so the ellipsoid $S_b^3$ reduces to $R^2\times S^1$ which is the large radius limit of $D^2\times S^1$. 
We can also demonstrate this explicitly from the metric \eqref{squashed 3 sphere metric in sph-coor}. 
When $\theta\to 0$, the metric \eqref{squashed 3 sphere metric in sph-coor} becomes
\begin{equation}
ds^2 = \tilde{l}^2 d\theta^2+l^2 d\varphi^2+\tilde{l}^2 \theta^2 d\chi^2 
= l^2 d\varphi^2+(d(\tilde{l}\theta)^2+(\tilde{l}\theta)^2 d\chi^2).
\end{equation}
The metric $d(\tilde{l}\theta)^2+(\tilde{l}\theta)^2 d\chi^2$ is exactly the metric of a plane in polar form, with $\tilde{l}\theta$ and $\chi$ identifying with the radial coordinate and the polar angle respectively. 
Notice that when $\theta \to 0$ and $\tilde{l} \to \infty$, we can keep $\tilde{l}\theta$ fixed and arbitrary, while $l$ is kept fixed and finite. We have therefore showed that the metric of ellipsoid $S_b^3$ indeed reduces to that of $R^2\times S^1$ in $b\to \infty$ limit. 

To localize the partition function on the Higgs branch, we need to introduce a deformation term. The path integral is insensitive to the choice of the deformation term, as long as the deformation term is a $\mathcal{Q}$-exact term. Since the Lagragians for vector multiplet and chiral multiplet are $\mathcal{Q}$-exact \eqref{lagragian}, we choose them to be our deformation term, 
\begin{align}
& \mathcal{L} \rightarrow \mathcal{L}+ t \delta_{\mathcal{Q}}V, \nonumber \\
& \delta_{\mathcal{Q}}V = \mathcal{L}_{v.m} +\mathcal{L}_{c.m}.
\end{align}
Note that our choice of deformation term does not include a FI term. Thus, the effective FI paramter of the deformed action is $\xi_{\rm eff}=\xi/(1+t)$. If we find out the saddle points of $\delta_{\mathcal{Q}}V$ after taking the $t \rightarrow \infty$ limit, the effective FI parameter vanishes and the solution leads to Coulomb branch. We can however take the limit other way round, i.e. performing the saddle point approximation at large but finite $t$ before taking $t$ to infinity \cite{Doroud:2012xw}. 
This requires us to find all saddle points of $\mathcal{L} \rightarrow \mathcal{L}+ t \delta_{\mathcal{Q}}V$.
The equation of motion for the D-term coming from the deformed action is 
\begin{equation}
D+ \frac{Q\sigma+m}{f}+ iQ\bar{\phi}\phi + \xi_{\rm eff}=0.
\end{equation}
It can be seen from \eqref{conf_1} and \eqref{conf_2} that 
\be
D+ \frac{Q\sigma+m}{f}=0, \quad
2iF_{23}-\frac{2}{3}\sigma V_1=0
\ee
for arbitrary $\theta$. We therefore deduce that
\begin{equation}
iQ\bar{\phi}\phi + \xi_{\rm eff}=0,
\end{equation}
which together with $(Q\sigma+m)\phi=0$, imply that the Coulomb branch is lifted and the path integral is localized to the Higgs branch instead. 
Now around $\theta=0$ and in the presence of FI parameter, the supersymmetric variations for the fermions reduce to 
\begin{align}
\label{vortex_eqn}
& D_2 \sigma = 2iF_{12}, \quad
 D_1 \sigma = 0, \quad
 D_3 \sigma = - 2 F_{31}, \nonumber \\
& iQ\bar{\phi}\phi + \xi_{\rm eff} -2iF_{23}=0, \nonumber \\
& D_2\phi - iD_3 \phi=0, \quad i D_1 \phi + \frac{1}{3}\phi (1-\frac{1}{b^2}) +\frac{\phi}{\tilde{l}}=0, \nonumber \\
& F=0, \quad  (Q\sigma+m)=0,
\end{align}
which are precisely the vortex equations.
Instead of $\theta=\pi$ on $S^2$, there is another vortex contribution at $\theta=\pi/2$, which corresponds to the vortex with fugacity $\q=\mathrm{exp}(2\pi i /b^2)$. The vortex equations at $\theta=\pi/2$ can be obtained as the way we obtained that at $\theta=0$, but it is interesting to note that we can get the equations by simply exchanging the subscripts 1 and 2 of the vortex equations at $\theta=0$, and performing the transformations,
\begin{equation}
iF_{23} \leftrightarrow -F_{31}, \quad b^2 \leftrightarrow \frac{1}{b^2}.
\end{equation}
In fact, this reflects the S-fusion of the boundaries of two solid tori.
Therefore, in localization computation we must integrate over the moduli space of solutions of vortices at $\theta=0$ and $\theta=\pi/2$.

For the vortex solution have finite energy, the scalars in the chiral multiplets must have the form at infinity,
\begin{equation}
\phi_a\sim e^{i f_a \theta},
\end{equation}
where $\theta$ is the angle parametrizing the circle at infinity, and the covariant derivative must vanish,
\begin{equation}
\left(\partial -i \sum_{b=1}^s (Q_u)_b^a A^b\right)\phi_a=0.
\end{equation}
For the covariant derivative to vanish, the gauge fields must have the following forms at infinity,
\begin{equation}
A^a \sim \hat{f}^a \partial \theta
\end{equation}
and satisfy the condition,
\begin{equation}
f^a=\sum_{b=1}^s (Q_u)^a_b \hat{f}^b, \quad a=1,2,...,s.
\end{equation}
As usual, $\hat{f}^a$ is the flux number of the background field, which is defined as
\begin{equation}
\hat{f}^a=\frac{1}{2\pi} \int d^2x F^a_2.
\end{equation}
If we have single gauge group and chiral multiplets of unit charges, flux is obviously equal to winding number which is an integer. However flux number $\hat{f}_a$ can be fractional in general and related to winding number $f_a$ via $\hat{f}_a = (Q_u^{-1})^b_a f_b$. In our subsequent vortex partition function computation, we restrict flux numbers to be integers, or more precisely, as explained in previous section, we are considering a specific vacuum admitting only integer vortices.

\subsection{Equivariant Localization}\label{EquivLoc}
\paragraph{}
For simplicity, we can send the radius of the disk $D^2$ to infinity, so that we place the theory in a background $R^2 \times_q S^1$. If we send the radius $\beta$ of $S^1$ to zero, the theory reduces to a two-dimensional $\mathcal{N}=(2,2)$ gauge theory on $R^2$ with an $\Omega$-deformation (with parameter $\epsilon$) which has been considered in \cite{Shadchin:2006yz, Yoshida:2011au, Fujimori:2012ab}. Since we want to study the vortex on $R^2 \times_q S^1$, we will keep $\beta$ finite. The resultant partition functions can now be interpreted as a K-theory generalization.

To check whether \eqref{HBlock2} the holomorphic block for Theory A coincides with a K-theory vortex partition function, we need to calculate the vortex partition function for $U(1)^s$ quiver gauge groups in principle. It is difficult to obtain this partition function by using equivariant localization, for our purpose here, we will restrict ourselves to check the simplest case with one $U(1)$ gauge group and $N$ fundamental chiral multiplets. However, one should note that the three dimensional mirror dual of such a theory has $U(1)^{N-1}$ quiver gauge groups, we therefore also verify the vortex partition function for this quiver theory. 

If we further restrict the charges of the chiral fields to be unity, the moduli space of the vortex equations \eqref{vortex_eqn} can be split into discrete components $\mathcal{M}_f$, where $f$ is the winding number for the non-zero valued scalar field $\phi$. 
This moduli space admits ADHM-like construction, so the vortex world volume theory can be reduced to a gauged matrix model which involves a $U(f)$ vector multiplet $(\varphi,\bar{\varphi},\lambda,\bar{\lambda},D)$. In addition, there is also an adjoint chiral multiplet $(B,\rho_{+})$ in our matrix model. The matter fields reduce to one chiral multiplet $(I,\mu_+)$ which is $f \times 1$ matrix and ${N-1}$ anti-chiral multiplets $(J_p,\nu_{p+})$ which is also a $f\times 1$ matrix. The plus and minus sign correspond to the left and right fermions, but only left fermion contributes because of the mass deformation.
The actions for the vortex matrix model are summarized in appendix \ref{actions for vortex}. 
The BRST transformations for the fields are:
\begin{align*}
& \mathcal{Q}_\epsilon \varphi=0,\quad \mathcal{Q}_\epsilon \bar{\varphi}=\eta, \quad \mathcal{Q}_\epsilon\eta=[\varphi,\bar{\varphi}],
\\& \mathcal{Q}_\epsilon D=[\varphi,\chi],\quad \mathcal{Q}_\epsilon\chi=D,
\\
& \mathcal{Q}_\epsilon I=\mu_+, \quad \mathcal{Q}_\epsilon\mu_+ = \varphi I+ I m_1,
\quad \mathcal{Q}_\epsilon I^\dagger =-\mu_{+}^\dagger, \quad \mathcal{Q}_\epsilon\mu_{+}^\dagger= I^\dagger  \varphi+m_1^* I^\dagger ,
\\
& \mathcal{Q}_\epsilon J_p=\nu_{p+}, \quad \mathcal{Q}_\epsilon\nu_{p+} = - J_p  \varphi-m_{d,p} J_p,
\quad \mathcal{Q}_\epsilon J_p^\dagger =\nu_{p+}^\dagger, \quad \mathcal{Q}_\epsilon\nu_{p+}^\dagger=-   \varphi J_p^\dagger- J_p^\dagger m_{d,p}^* ,
\\
& \mathcal{Q}_\epsilon B=\rho_+, \quad \mathcal{Q}_\epsilon\rho_+=[\varphi,B]-\epsilon B,
\quad \mathcal{Q}_\epsilon B^\dagger= -\rho_+^{\dagger}, \quad \mathcal{Q}_\epsilon \rho_+^{\dagger}=[\varphi,B^\dagger]-\epsilon B^\dagger.
\end{align*}
Here the vector $m_u$ in \eqref{matrix model matter part} reduces to $m_1$ here as we are considering single $U(1)$ gauge group and have selected a specific vacuum.

Let us now consider the localization method. Firstly, we introduce the vector field $\mathcal{Q}^*$ which acts on the fields and generates the BRST transformation
\begin{align*}
\mathcal{Q}^* &= \bigg([\varphi,\chi]\frac{\partial}{\partial D}+ \eta \frac{\partial}{\partial \bar{\varphi}}+ \rho_+ \frac{\partial}{\partial B} + D \frac{\partial}{\partial \chi} + [\varphi,\bar{\varphi}]\frac{\partial}{\partial \eta} + ([\varphi,B]-\epsilon B) \frac{\partial}{\partial \rho_+}\bigg)
\\
 + &  \Big( \varphi I+ I m_1 \Big) \frac{\partial}{\partial \mu_{+}} + \mu_{+} \frac{\partial}{\partial I} + \bigg( \Big(- J \varphi-m_{d} J\Big) \frac{\partial}{\partial \nu_{+}} +\nu_{+} \frac{\partial}{\partial J} \bigg).
\end{align*}
The critical points, $\mathcal{Q}^*=0$, are then given by 
\begin{align}
\label{Fixed_Point_1}
\varphi I+ I m_1 &=0 , \\
\label{Fixed_Point_2}
- J  \varphi-m_{d} J &=0, \\
\label{Fixed_Point_3}
[\varphi,B]-\epsilon B &=0.
\end{align}
These equations are similar to the equations involved in the ADHM construction of instanton and are already solved in \cite{Nekrasov:2002qd}. The solutions are
\begin{equation}
\label{Fixed_Point_Sol}
\varphi^\omega =-m_1 + (\omega-1) \epsilon,\quad \omega=1,...,f;  \quad
J=0.
\end{equation}
Instead of doing contour integral around these fixed points, we introduce the following vector space,
\begin{align*}
&B_{\mathbb{C}}\in \mathrm{Hom}_\mathbb{C}(\hat{V},\hat{V}), \quad \mathrm{dim}_\mathbb{C}(\hat{V})=f ,\\
&  I_{\mathbb{C}}\in \mathrm{Hom}_\mathbb{C}(G,\hat{V}), \quad \mathrm{dim}_\mathbb{C}(G)=1, \\
&J_{p\mathbb{C}}\in \mathrm{Hom}_\mathbb{C}(\hat{V},H), \quad \mathrm{dim}_\mathbb{C}(H)=N-1.
\end{align*}
Define a torus action $U_\epsilon(1)\times U(1) \times U(1)^{N-1}$ on $\mathcal{M}_f$, such that
\begin{equation}
U_\epsilon(1):\quad (B_{\mathbb{C}}, I_{\mathbb{C}},J_{\mathbb{C}}) \to (q^{-1}B_{\mathbb{C}},I_{\mathbb{C}},J_{\mathbb{C}}),
\end{equation}
where $q=e^{-2\pi\beta\epsilon} (\epsilon \in \mathbb{R}^+)$.  
According to the action of $U_\epsilon(1)$, we modify the $B_{\mathbb{C}} \in \mathrm{Hom}_\mathbb{c}(\hat{V},\hat{V})\otimes Q$, where $Q$ is the one dimensional space on which the $U_\epsilon(1)$ acts. In addition, the action $U_\epsilon(1)\times U(1) \times U(1)^{N-1}$ on $(B_{\mathbb{C}},I_\mathbb{C},J_\mathbb{C})$ is
\begin{align}
U(1)_1 &: \quad (B_{\mathbb{C}}, I_{\mathbb{C}},J_{\mathbb{C}}) 
\to (B_{\mathbb{C}}, I_{\mathbb{C}}Q_{m_1}^{-1}, J_{\mathbb{C}}), \nonumber \\
U(1)^{N-1} &: \quad (B_{\mathbb{C}}, I_{\mathbb{C}},J_{\mathbb{C}})  
 \to (B_{\mathbb{C}}, I_{\mathbb{C}},Q_{m_d} J_{\mathbb{C}}), \nonumber
\end{align}
where
\begin{equation}
Q_{m_1}=e^{2\pi\beta m_1}, \quad
Q_{m_d}=\mathrm{diag}(e^{2\pi\beta m_{d,2}},e^{2\pi\beta m_{d,3}},...,e^{2\pi\beta m_{d,N}}).
\end{equation}
Furthermore, we define the action $g \in \mathrm{Hom}_\mathbb{C} \big(U_\epsilon(1)\times U(1) \times U(1)^{N-1}, U(f)\big)$ by
\begin{align*}
g: \quad &(B_{\mathbb{C}}, I_{\mathbb{C}},J_{\mathbb{C}}) 
 \to (g(t)B_{\mathbb{C}} g(t)^{-1},~g(t)I_{\mathbb{C}},~J_{\mathbb{C}}g(t)^{-1})\\
& t=(q,Q_{m_1},Q_{m_d}) \in U_\epsilon(1)\times U(1) \times U(1)^{N-1}
\end{align*}
and set the $U(f)$ gauge transformation as
\begin{equation}
\label{Gauge_Trans_1}
g(t) = \mathrm{diag} (e^{2\pi\beta \varphi^1},e^{2\pi\beta \varphi^2},...,e^{2\pi\beta \varphi^{f}}).
\end{equation}
Applying the fixed point conditions,
\begin{align}
g(t)B_{\mathbb{C}} g(t)^{-1} &= q^{-1}B_{\mathbb{C}}, \nonumber \\
g(t) I_{\mathbb{C}} &= I_{\mathbb{C}} Q_{m_1}^{-1}, \nonumber \\
J_{\mathbb{C}}g(t)^{-1} &= Q_{m_d} J_{\mathbb{C}},
\end{align}
we can get the equations \eqref{Fixed_Point_1}, \eqref{Fixed_Point_2} and \eqref{Fixed_Point_3} as the infinitesimal forms.
Given the solutions \eqref{Fixed_Point_Sol}, we can decompose the space $\hat{V}$ as:
\begin{align*}
\hat{V} &= \bigoplus_\omega \hat{V}(\omega), \quad \text{where} \quad \hat{V}(\omega) := \{v \in \hat{V} | g(t)v =  q^\omega Q_{m_1}^{-1} \}.
\end{align*}

\subsection{Tangent Space of the Fixed Point and Vortex Partition Function}
\paragraph{}
The tangent space of $\mathcal{M}$ at the fixed points is 
\begin{align*}
T_x \mathcal{M} &= \bigg(\mathrm{Hom}_\mathbb{C}(\hat{V},\hat{V}\otimes Q) \oplus \mathrm{Hom}_\mathbb{C}(G,\hat{V})  \oplus \mathrm{Hom}_\mathbb{C}(\hat{V},H) \bigg) /\mathrm{Hom}_\mathbb{C}(V,V).
\end{align*}
We will use the same symbols for the characters and the representation spaces. The characters are
\begin{align*}
\hat{V} &= \sum_{\omega=1}^{f} q^{\omega-1} Q_{m_1}, \quad Q= q^{-1}, \quad G = Q_{m_u}, \quad H = \sum_{p=2}^{N-1} Q_{m_{d,p}}. 
\end{align*}
The character of the moduli space is
\begin{equation}
{\rm ch}(T_x \mathcal{M}) =(\hat{V}^* \times Q - \hat{V}^*)\times V +(G^* \times  \hat{V}) + (\hat{V}^* \times H ) .
\end{equation}
Thus the character of $\mathcal{M}$ is given by
\begin{eqnarray}
 {\rm ch}(T_x \mathcal{M})  &=&  \left( \sum_{\tilde{\omega}=1}^{\tilde{f}} (q^{-1})^{\tilde{\omega}-1} Q_{m_1}^{-1} q^{-1}-  \sum_{\tilde{\omega}=1}^{\tilde{f}} (q^{-1})^{\tilde{\omega}-1} Q_{m_1}^{-1} \right) \times \sum_{\omega=1}^{f} q^{\omega-1}  Q_{m_1}  +   \left( Q_{m_1}^{-1}  \times  \sum_{\omega=1}^{f} q^{\omega-1} Q_{m_1} \right)\nn \\
&+& \Big( \sum_{\omega=1}^{f} (q^{-1})^{(\omega-1)}  e^{-2\pi\beta m_1}  \times \sum_{p=2}^{N} e^{2\pi\beta m_{d,p}} \Big) \nonumber
\\
&= & \Big(\sum_{\omega=1}^{f} q^{\omega-\tilde{f}-1} \Big) + \sum_{p=2}^{N} \Big( e^{2\pi\beta(m_{d,p}-m_1)}\sum_{\omega=1}^{f} (q^{-1})^{(\omega-1)} \Big) .
\end{eqnarray}
Therefore the vortex partition function on $R^2\times S^1$ with $N$ fundamental chiral multiplets is
\begin{equation}\label{vortex partition function}
Z_{\rm vortex}
= \sum_{f=0}^\infty \frac{z^{f}}{(q^{-1}; q^{-1})_f} \prod\limits_{p=2}^N  \frac{1}{  \left(e^{2\pi\beta(m_{d,p}-m_1)};q^{-1}  \right)_f}.
\end{equation}
If we identify $z$ as ${\bf y}_a^{-1}$ with $a=1$, $Q_u=1$ and $\beta=b$, this vortex partition function is indeed what we identify as the vortex part of the holomorphic block given in \eqref{HBlock2} in this simple setting, i.e. one $U(1)$ gauge group and $N$ chiral multiplets.
By matching the two partition functions, we conclude that the holomorphic block obtained from factorizing the ellipsoid partition function indeed admits a K-theory vortex partition function interpretation. 

The ADHM-like construction for arbitrary charges case seems difficult and is currently unavailable to the best of our knowledge. One reason is that the number of moduli for vortex solutions increase to $\sum_i^N Q_i \hat{f}$ in complex dimension. Here we would like to propose possible modifications to the previous calculation to obtain the partition function for $U(1)$ gauge group with $N$ chiral multiplets of arbitrary charges. Since each chiral multiplet has distinct charge, the matrix $J$ splits into $J_p,~p=2,...,N$. Moreover, we claim that the $J_p$ is a $1 \times \frac{Q_p}{Q_1}{f}$ matrix (we still restrict $\frac{Q_p}{Q_1}=\text{integer}$). Then, the $U(f)$ group will act on $J_p$'s in different representations. The fixed point equations remain to be \eqref{Fixed_Point_1},  \eqref{Fixed_Point_3}, but \eqref{Fixed_Point_2} becomes
\begin{align*}
-J_p \varphi_p -m_{d,p}J_p=0.
\end{align*}
The solution is still \eqref{Fixed_Point_Sol}. However, we need that 
\begin{equation*}
\varphi_p=\mathrm{diag}\left(-\frac{Q_p}{Q_1}m_1,~-\frac{Q_p}{Q_1}m_1+\epsilon,.....,~-\frac{Q_p}{Q_1}m_1+ (\frac{Q_p}{Q_1}f-1)\epsilon\right).
\end{equation*}
Then the character of the moduli space is
\begin{equation}
{\rm ch}(T_x \mathcal{M}) =(\hat{V}^* \times Q - \hat{V}^*)\times V +(G^* \times  \hat{V}) + \sum_{p=2}^N (\hat{V}^{p*} \times H_p ) 
=  \Big(\sum_{\omega=1}^{f} q^{\omega-\tilde{f}-1} \Big) + \sum_{p=2}^{N} \Big( e^{2\pi\beta(m_{d,p}-\frac{Q_p}{Q_1}m_1)}\sum_{\omega=1}^{\frac{Q_p}{Q_u} f} (q^{-1})^{(\omega-1)} \Big).
\end{equation}
Hence, the partition function is
\begin{align}
\label{VortexPart2}
Z_{\rm vortex}
= \sum_{\{f=0\}}^\infty 
&\frac{z'^{f}}{ \prod\limits_{\omega=1}^{f} (1-q^{-\omega})} \prod\limits_{p=2}^N  \frac{1}{ \prod\limits_{\omega=1}^{\frac{Q_p}{Q_1}f} \bigg(1-e^{2\pi\beta(m_{d,p}-\frac{Q_p}{Q_1}m_1)} (q^{-1})^{\omega-1} \bigg)} \quad _.
\end{align}
Equation \eqref{VortexPart2} can be regarded as  the generalization of vortex partition function \eqref{vortex partition function}. 
In particular, if we now identify $z'$ with $y_a^{-1/Q_1}$ or ${\bf y}^{-1}_a$ with $a=1$ in \eqref{HBlock2} and $\beta=b$, moreover dividing the non-negative integers $\{f\}$ as in \eqref{FactorPole} into $|Q_1|$ different sectors, yielding $|Q_1|$ different sets of integer (i. e. $f=Q_1\hat{f}$) and fractional vortices (otherwise), the vortex part in holomorphic block \eqref{HBlock2} precisely corresponds to the integer vortex sector, and the remaining contributions in \eqref{VortexPart2} should match with the remaining holomorphic blocks associated with the vacua that admit fractional vortices.

\acknowledgments
This work was supported in part by National Science Council through the grant No.101-2112-M-002-027-MY3, Center for Theoretical Sciences at National Taiwan University and Kenda Foundation. Heng-Yu Chen also acknowledges the stimulating discussions with Toshiaki Fujimori, Kazuo Hosomchi, Maurizio Romo and Daisuke Yokoyama and the hospitality of Yukawa Institute for Theoretical Physics, Kyoto and Kavli Institute for the Physics and the Mathematics of the Universe (KIPMU), University of Tokyo, where parts of this work were completed.

\appendix

\section{Jacobian Factors Between Two Mirror Theories}
\label{determinant}
\paragraph{}
Here we would like to show that the ratio $|\det (Q_u)|/|\det (\hat{Q}_d)|$ is invariant under changing the choice of $\{Q_u\}$ and $\{\hat{Q}_d\}$.
Define charges of two mirror pair as:
\begin{eqnarray}
Q=\{A_a, B_a, C_a, \cdots M_a, N_a, O_a \cdots\},\nn\\
\hat{Q}=\{\alpha_p, \beta_p, \gamma_p,\cdots, \xi_p, \eta_p, \zeta_p, \cdots \},
\end{eqnarray}
and a certain choice:
\begin{eqnarray}
Q_u=\{A_a, B_a, C_a, \cdots P_a\},&\quad & \{M_a, N_a,O_a,\cdots P'_a\}\notin Q_u\nn
\\
\hat{Q}_d=\{\xi_p, \eta_p, \zeta_p, \cdots \pi_p\},&\quad & \{\a_p,\b_p,\gamma_p, \cdots\pi'_p\}\notin \Q_d
\end{eqnarray}
where $A_a, B_a, C_a, \cdots M_a. N_a, O_a. P_a$ and $\alpha_p, \beta_p, \gamma_p, \xi_p, \eta_p, \zeta_p, \cdots \pi_p$ are $s$-column vectors and $N-s$-column vectors respectively.

Now, without losing generality, we make the other choice by changing $P_a\rightarrow P'_a$ and $\pi_p\rightarrow \pi'_p$, getting:
\begin{eqnarray}
Q'_u=\{A_a, B_a, C_a, \cdots P'_a\}\nn,
\\
\hat{Q}_d'=\{\xi_p, \eta_p, \zeta_p, \cdots \pi'_p\}.
\end{eqnarray}
They will satisfy the orthogonal condition:

\begin{equation}
\label{determinatant orthogonal}
A_a\alpha_p+B_a\beta_p+\cdots+M_a\xi_p+N_a\eta_p+\cdots+P_a\pi'_p+P'_a\pi_p=0,
\end{equation}
Assume the two ratio are non-trivially related by a number $R(\{Q_u\},\{Q'_u\})$ i.e.
\begin{equation}
\frac{\det Q_u}{\det \hat{Q}_d}=R\frac{\det Q_u'}{\det \hat{Q}_d'}
\Rightarrow
\frac{\epsilon_{ijk\cdots}A_iB_j\cdots P_k}
{\epsilon_{pqr\cdots}\xi_p\eta_q\cdots\pi_r}
=
R
\frac{\epsilon_{ijk\cdots}A_iB_j\cdots P'_k}{\epsilon_{pqr\cdots}\xi_p\eta_q\cdots\pi'_r}.
\end{equation}
\begin{equation}
\Rightarrow\epsilon_{ijk}\epsilon_{pqr}A_iB_j\xi_p\eta_p\cdots(P_k\pi'_r-RP'_k\pi_r)=0.
\end{equation}
Using the orthogonal condition \eqref{determinatant orthogonal}, we get:
\begin{equation}
\epsilon_{ijk}\epsilon_{pqr}A_iB_j\xi_p\eta_p\cdots P'_k\pi_r(-1-R)=0 \Rightarrow R=-1.
\end{equation}
This reflects the fact that $|\det{Q_u}|/|\det{\hat{Q}_d}|$ is a constant independent from the choice of $\{\{Q_u\},\{Q'_u\}\}$.

\section{Killing Spinors and SUSY Theories on Ellipsoid}
\label{SqS3SUSY}
\paragraph{}
We need to choose a spinor which satisfies the Killing spinor equation to parametrize infinitesimal supersymmetry transformations. For the coordinate and metric \eqref{squashed 3 sphere metric in sph-coor} ,the Killing spinor equation  for a spinor field $\psi$ reduces to
\begin{align}
\label{killing spinor eqn}
if\partial_\varphi \psi &= l \gamma^2(\frac{1}{2}\mathrm{sin}\theta+\mathrm{cos}\theta \partial_\theta)\psi , \nonumber \\
if\partial_\chi \psi &= \tilde{l}\gamma^1(\frac{1}{2}\mathrm{cos}\theta+\mathrm{sin}\theta \partial_\theta)\psi .
\end{align}
If $l=\tilde{l}$, the general solution of equation \eqref{killing spinor eqn} is 
\begin{equation}
\psi_{st}=
\begin{pmatrix}
e^{\frac{i}{2} (s\chi+t\varphi-st\theta)}\\
-se^{\frac{i}{2} (s\chi+t\varphi+st\theta)}
\end{pmatrix},\quad s,t=\pm
\end{equation}
satisfying $D_\mu \psi_{st}=-\frac{ist}{2l}\gamma_\mu \psi_{st}$.
However, when $l \neq \tilde{l}$, the spinor fields $\psi_{st}$ fail to satisfy the Killing spinor equations,
\begin{equation}
\label{Killing}
-\frac{ist}{2l}\gamma_\mu \psi_{st}=D_\mu \psi_{st}-iV_\mu^{(st)}\psi_{st},
\end{equation}
where
\begin{equation}
V^{(st)}=\frac{t}{2}(1-\frac{l}{f})d\varphi + \frac{s}{2}(1-\frac{\tilde{l}}{f})d\chi.
\end{equation}
One can reinterpret the unwanted term in the right hand side of \eqref{Killing} as the coupling to a background $U(1)$ gauge field. We choose to turn on the gauge field $V=V^{+-}$ so that the spinors $\epsilon=\psi_{+-}$ and $\bar{\epsilon}=\psi_{-+}$ satisfy the Killing spinor equation with $U(1)$ charges $\pm1$.

In order to derive the SUSY saddle point equations, we need to impose the condition $\delta(fermion)=0$. Therefore, we only write down the SUSY transformations of 3d $\mathcal{N}=2$ gauge theories on squashed sphere for the fermions. For other details, see \cite{Hama:2011ea}.
The SUSY transformations for the fermions in a vector multiplet are
\begin{align}
\label{SUSY trans for vector fermion}
\delta\lambda &=\frac{1}{2}\gamma^{\mu\nu}\epsilon F_{\mu\nu}-D\epsilon+i\gamma^\mu\epsilon D_\mu \sigma+\frac{2i}{3}\sigma\gamma^\mu D_\mu \epsilon, \nonumber \\
\delta\bar{\lambda} &=\frac{1}{2}\gamma^{\mu\nu}\bar{\epsilon}F_{\mu\nu}+D\bar{\epsilon}-i\gamma^\mu\bar{\epsilon} D_\mu \sigma-\frac{2i}{3}\sigma\gamma^\mu D_\mu \bar{\epsilon}.
\end{align}
The SUSY transformations for the fermions in a chiral multiplet with charge $Q$ are
\begin{align}
\label{SUSY trans for chiral fermion}
\delta\psi &= i\gamma^\mu \epsilon D_\mu \phi + iQ\epsilon \sigma \phi +\frac{2qi}{3}\gamma^\mu D_\mu \epsilon \phi + \bar{\epsilon}F, \nonumber \\
 \delta\bar{\psi} &= i\gamma^\mu \bar{\epsilon} D_\mu \bar{\phi} + iQ\bar{\phi} \sigma \bar{\epsilon} +\frac{2qi}{3}\bar{\phi}\gamma^\mu D_\mu \bar{\epsilon} + \bar{F}\epsilon.
\end{align}
Please note that we have turn on the background $U(1)$ gauge field, so $D_\mu$ denotes the covariant derivative with respect to gauge, local lorentz and background $U(1)$ gauge field.

The supersymmetric lagragians for vector multiplet,chiral multiplet and FI term are \cite{Hama:2011ea},
\begin{align}
\mathcal{L}_{v.m} &= \mathrm{Tr} \bigg( \frac{1}{4} F_{\mu\nu}F^{\mu\nu} + \frac{1}{2} D_\mu \sigma D^\nu \sigma + \frac{1}{2} (D+\frac{\sigma}{f})^2 +\frac{i}{2}\bar{\lambda} \gamma^\mu D_\mu \lambda +\frac{i}{2}\bar{\lambda}[\sigma,\lambda]+\frac{1}{4f}\bar{\lambda} \lambda \bigg) \nonumber \\
\mathcal{L}_{c.m} &=D_\mu\bar{\phi}D^\mu \phi + Q^2\bar{\phi} \sigma^2 \phi + \frac{i(2q-1)}{f}Q\bar{\phi} \sigma \phi - \frac{q(2q-1)}{2f^2}\bar{\phi}\phi +\frac{q}{4}R \bar{\phi}\phi + iQ\bar{\phi}D\phi +\bar{F}F \nonumber \\
&-i\bar{\psi}\gamma^\mu D_\mu \psi +i Q\bar{\psi}\sigma\psi -\frac{(2q-1)}{2f}\bar{\psi}\psi +iQ\bar{\psi}\lambda\phi -iQ\bar{\phi}\bar{\lambda}\psi \nonumber \\
\mathcal{L}_{FI}&=D-\frac{\sigma}{f}.
\end{align}
The lagragians can be written into $\mathcal{Q}$-exact forms,
\begin{align}
\label{lagragian}
\bar{\epsilon}\epsilon \mathcal{L}_{v.m}&=\delta_{\bar{\epsilon}}\delta_\epsilon \bigg(\bar{\psi}\psi- 2i\bar{\phi}\sigma\phi+\frac{(2q-1}{f}\bar{\phi}\phi \bigg) \nonumber \\
\bar{\epsilon}\epsilon \mathcal{L}_{c.m}&= \delta_{\bar{\epsilon}}\delta_\epsilon \mathrm{Tr}\bigg(\frac{1}{2}\bar{\lambda}\lambda-2D\sigma \bigg).
\end{align}

\section{Supersymmetric Actions for the Vortex Matrix Model}
\label{actions for vortex}
\paragraph{}
The actions for the K-theoretic vortex are
\begin{align}
S_G &= \frac{1}{2}[\varphi,\bar{\varphi}]^2 + D^2 + g^2 \zeta D + 2 \bar{\lambda}_{-}[\varphi,\lambda_{-}],
\\
\label{matrix model matter part}
S_m &= -2  I^\dagger (\bar{\varphi}+m_u^*) (\varphi+m_u) I +  I^\dagger D I  - \sqrt{2}\mu_{+}^\dagger \bar{\varphi} \mu_{+} + i\sqrt{2}[I^\dagger \lambda_{-}\mu_{+}-\mu_{+}^\dagger \bar{\lambda}_{-} I],
\\
S_m' &= \sum_{p=2}^N \bigg(-2  J_p (   \bar{\varphi}+m_{d,p}^*) (  \varphi+m_{d,p}) J_p^\dagger +  J_p D J_p^\dagger - \sqrt{2}\nu_{p+}^\dagger \bar{\varphi} \nu_{p+} + i\sqrt{2} [J_p^\dagger \lambda_{-}\nu_{p+}-\nu_{p+}^\dagger \bar{\lambda}_{-} J_p]\bigg),
\\
S_A &= \mathrm{Tr}\bigg( |[ \varphi,B^\dagger]|^2 +  D[B,B^\dagger]- \sqrt{2} [\bar{\varphi},\rho_+^{\dagger}]\rho_+ i\sqrt{2} \big( B[\rho_+^{\dagger},\bar{\lambda}_{-}] + B^\dagger [\lambda_{-},\rho_+] \big) \bigg).
\end{align}

\end{document}